\documentclass[12pt]{article}

\usepackage[left=2.8cm,right=2.8cm,top=2.8cm,bottom=2.8cm]{geometry}

\usepackage[colorlinks=true,linkcolor=black,citecolor=black,urlcolor=black,
filecolor=black]{hyperref}

\usepackage{latexsym}
\usepackage{amsmath}
\usepackage{amsfonts}
\usepackage{mathrsfs}
\usepackage{dsfont}
\usepackage{verbatim}
\usepackage[nosort]{cite}

 \csname
@addtoreset\endcsname{equation}{section}

\def\a{\alpha}
\def\b{\beta}

\def\d{\delta}

\def\e{\epsilon}

\def\h{\eta}
\def\th{\theta}

\def\k{\kappa}
\def\l{\lambda}
\def\L{\Lambda}
\def\m{\mu}
\def\n{\nu}
\def\x{\xi}

\def\p{\pi}

\def\r{\rho}

\def\s{\sigma}

\def\vf{\varphi}

\def\o{\omega}

\def\cA{{\cal A}}
\def\cB{{\cal B}}
\def\cC{{\cal C}}
\def\cD{{\cal D}}
\def\cE{{\cal E}}
\def\cF{{\cal F}}

\def\cL{{\cal L}}

\def\cO{{\cal O}}

\def\cT{{\cal T}}

\def\cW{{\cal W}}

\def\be{\begin{equation}}
\def\ee{\end{equation}}
\def\bea{\begin{eqnarray}}
\def\eea{\end{eqnarray}}
\def\ba{\begin{array}}
\def\ea{\end{array}}

\def\nn{\nonumber}

\def\tr{\text{tr}}

\newcommand{\os}[2]{\overset{\scriptscriptstyle (#1)}{#2}}

\def\ww{\wedge}

\def\comma{\,,\,}

\def\12{\frac{1}{2}}
\def\pr{\partial}

\begin{document}

\begin{flushright}
\begin{tabular}{c}
AEI-2012-081
\end{tabular}
\end{flushright}

\vspace{20pt}

\begin{center}


{\Large\sc Towards metric-like higher-spin \\[10pt]
gauge theories in three dimensions}


\vspace{25pt}
{\sc A.~Campoleoni, S.~Fredenhagen, S.~Pfenninger and S.~Theisen}

\vspace{10pt}
{\sl\small
Max-Planck-Institut f{\"u}r Gravitationsphysik\\
Albert-Einstein-Institut\\
Am M{\"u}hlenberg 1\\
14476 Golm,\ GERMANY\\[5pt]

\vspace{10pt}

{\it andrea.campoleoni@aei.mpg.de, stefan.fredenhagen@aei.mpg.de,\\ 
stefan.pfenninger@aei.mpg.de, stefan.theisen@aei.mpg.de} 
}

\vspace{60pt} {\sc\large Abstract}\end{center}

We consider the coupling of a symmetric spin-3 gauge
field $\varphi_{\mu\nu\rho}$ to
three-dimensional gravity in a second order metric-like formulation. 
The action that corresponds to an
$SL(3,\mathbb{R})\times SL(3,\mathbb{R})$ Chern-Simons theory in the
frame-like formulation is identified to quadratic order in
the spin-3 field. We apply our result to compute corrections to the
area law for higher-spin black holes using Wald's entropy formula.

\newpage


\tableofcontents

\section{Introduction and overview}\label{sec:intro}

By now much is known about the structure of
interacting field theories involving particles of spin greater than two. In
particular, Vasiliev proposed a set of
non-linear equations of motion that describe the interactions of an infinite
tower of gauge fields of increasing spin on (A)dS backgrounds \cite{Vasiliev}
(see \cite{rev-vas} for a review). This result rests upon a
description of the dynamics that mimics the frame approach to gravity: the
degrees of freedom are encoded in a set of differential forms taking values in
an infinite-dimensional extension of the Lorentz algebra, and the field
equations are manifestly invariant under diffeomorphisms. Alternative approaches to higher-spin interactions are also actively investigated in order to look for generalisations of Vasiliev's construction or to make more transparent various features of the interactions (see e.g.\ \cite{rev-metric1,rev-metric2,rev-metric3} for reviews). For instance, one can follow the path
of the metric formulation of gravity and encode the degrees of freedom of
a spin-$s$ particle in a symmetric tensor of rank $s$. The advantage with
respect to Vasiliev's strategy is the simplification of the field
content; the price to pay is, at present, the lack of an organising principle for the
non-linearities required by a consistent theory. To unravel this puzzle one can begin by building perturbatively the first interaction vertices; this has led, for instance, to a classification of cubic vertices for arbitrary massless particles in both Minkowski and (A)dS backgrounds of dimension $D \geq 4$ \cite{old-LC,Metsaev,MMR,ST,JLT}.\footnote{The classification of cubic interactions for arbitrary fields is discussed in a frame-like language in
\cite{vertices-frame}.} On the other hand, a complete metric-like reformulation of Vasiliev's
equations is not known, while the existence of other models that are
consistent beyond the cubic order is still controversial (see e.g.\
\cite{nogo-spin3,Polyakov,Taronna,vertices-frame,Tsulaia}).

In spite of closely related goals, the frame- and metric-like
formulations have evolved rather independently. For few
exceptions see e.g.\ \cite{parent1,HS-symm,parent2} and refs.\
therein. With both approaches having their own advantages and
drawbacks, an exchange of ideas is nonetheless expected to shed light
on both sides. The goal of this paper is to start to establish a firm
connection between them in three space-time dimensions, where
higher-spin gauge theories take a remarkably simple form compared to
their higher-dimensional counterparts. We focus on the gravitational
coupling of a symmetric tensor of rank 3.  In the frame-like language
this is described by a $SL(3,\mathbb{R}) \times SL(3,\mathbb{R})$
Chern-Simons (CS) theory when a negative cosmological constant is
present (see e.g.\ \cite{spin3} and the previous
works\cite{Blencowe1,Blencowe2}).  In appendix~C we will add a few
comments on the generalisation to $SL(N,\mathbb{R}) \times
SL(N,\mathbb{R})$ CS theories, which contain fields of spin
$2,3,\ldots,N$.

The frame-like theory is well understood, 
with and without cosmological constant:
one has to complement the gravity dreibein and
spin connection with two one-forms which play a similar role for the spin-3 field. 
The gauge connections can then be packed into two $sl(3,\mathbb{R})$-valued forms
($\cA=1,\dots,8$ and $a,b=0,1,2$)
\begin{subequations} \label{forms}
\begin{alignat}{5}
e & \,=\, e_\m{}^{\!\cA}\, J_\cA\, dx^\m & & =\, \left(\, e_\m{}^a J_a +
e_\m{}^{ab}\, T_{ab}\, \right) dx^\m \, , \\[10pt]
\o & \,=\, \o_\m{}^{\!\cA}\, J_\cA\, dx^\m & & =\, \left(\, \o_\m{}^a J_a +
\o_\m{}^{ab}\, T_{ab}\, \right) dx^\m \, ,
\end{alignat}
\end{subequations}
where $J_\cA$ denotes the full set of $sl(3,\mathbb{R})$ generators. The gravity dreibein 
$e_\mu{}^a$ and spin connection $\omega_\mu{}^a$ are associated with 
the generators $J_a$ of the principally embedded $so(2,1)\simeq sl(2,\mathbb{R})\hookrightarrow
sl(3,\mathbb{R})$. The remaining five generators $T_{ab}$ (with $T_{[ab]}=\eta^{ab}\,T_{ab}=0$) 
are associated to the spin-3 ``vielbein'' and
``spin connection''. One can then consider the action\footnote{For
$\ell^2>0$ (corresponding to a negative cosmological constant) one can
rewrite \eqref{CS} as the difference of two $sl(3,\mathbb{R})$ CS
actions. A cosmological constant is however not necessary in $D=3$,
and for $\ell^2\leq 0$ one can interpret \eqref{CS} as a CS action as
well (see e.g.\ \cite{review_3D} for more details).}
\begin{equation} \label{CS}
I \,=\, \frac{1}{16\p G} \int {\rm tr} \left(\, e \ww R \,+\, \frac{1}{3 \ell^2}\, e \ww e \ww e
\,\right) , \quad \textrm{with} \quad R \,=\, d \o + \o \ww \o \, .
\end{equation}
The trace is in the fundamental of $sl(3,\mathbb{R})$, $G$
is Newton's constant and $\ell$ the AdS radius. 

A first step towards the identification of the metric-like counterpart of
\eqref{CS} was taken in \cite{spin3}, where the metric and the spin-3 field were
expressed in terms of the connection one-forms \eqref{forms} as
\begin{equation} \label{fields}
g \,=\, \12\, \tr \left(\, e_\m e_\n \,\right) dx^\m dx^\n \, , \qquad \vf
\,=\,\frac{1}{6}\, \tr \left(\, e_\m e_\n e_\r \,\right) dx^\m dx^\n
dx^\r \, .
\end{equation}
The justification for \eqref{fields} is that the action \eqref{CS} is invariant under the transformations
\begin{subequations} \label{gauge_fields}
\begin{align}
& \d e \,=\, d\x \,+\, [\, \o \,,\, \x \,] \,+\, [\, e \,,\, \L \,] \, ,
\label{gauge_viel} \\
& \d \o \,=\, d\L \,+\, [\, \o \,,\, \L \,] \,+\, \frac{1}{\ell^2}\, [\, e \,,\, \x
\,] \, , \label{gauge_spin}
\end{align}
\end{subequations}
generated by $sl(3,\mathbb{R})$-valued parameters $\x$ and $\L$.
Those generated by $\L$ include and generalise local Lorentz transformations: therefore metric-like fields
should be invariant under them, and this is guaranteed by
\eqref{fields}
 (see also \cite{Wlambda} for a discussion of the $sl(N,\mathbb{R})$ case).
The transformations with parameters $\xi$ should then
give rise to the transformations of $g$ and $\vf$ under diffeomorphisms and a suitable deformation of the
linearised Fronsdal gauge symmetry \cite{Fronsdal}.

In the present paper we verify these statements up to the quadratic order in the spin-3 field, while keeping
all non-linearities in the metric. In particular, we show that the action \eqref{CS} can 
be rewritten in terms of the metric-like fields \eqref{fields} as
\begin{equation} \label{action}
\begin{split}
I\,& =\int \frac{d^3x\, \sqrt{-g}}{16\p G} \ \bigg\{ \left( R \,+\,
\frac{2}{\ell^2} \right) \,+\, \vf^{\,\m\n\r} \left( \cF_{\m\n\r}
\,-\, \frac{3}{2} \, g_{(\m\n}\, \cF_{\r)} \right) \,-\, \frac{3}{2}\, R\,
\vf_{\m\n\r}\, \vf^{\,\m\n\r} \\[10pt]
& + \frac{9}{4}\, R_{\rho\sigma} \Big(\,
2\,  \vf^{\,\rho}{}_{\m\n}\, \vf^{\,\sigma\m\n}
-\, \vf^{\,\rho}{}\, \vf^{\,\sigma}{} \,\Big)
- \frac{1}{\ell^2}\, \Big(\, 6\, \vf_{\m\n\r}\, \vf^{\,\m\n\r}
-\, 9\, \vf_{\m}\, \vf^{\,\m} \,\Big) \bigg\} +\,
\cO\!\left(\vf^4\right) ,
\end{split}
\end{equation}
where $R_{\m\n}$ is the Ricci tensor, $\nabla$ is the
Levi-Civita connection for the metric defined in \eqref{fields} and $\cF_{\m\n\r}$ denotes the
covariantised Fronsdal tensor
\begin{equation}
\cF_{\m\n\r} =\, \Box\, \vf_{\m\n\r} \,-\, \frac{3}{2} \left(\, \nabla^\l
\nabla_{\!(\m\,} \vf_{\n\r)\l} +  \nabla_{\!(\m} \nabla^{\l\,} \vf_{\n\r)\l}
\,\right) \,+\, 3\, \nabla_{\!(\m}\nabla_{\!\n}\, \vf_{\r)}\,.
\end{equation}
We also have defined $\vf_\mu \equiv \vf_{\mu\l}{}^\l$ and, likewise, $\cF_\m$ 
is the trace of the Fronsdal tensor. 
Indices between parentheses are meant to be
symmetrised, and dividing by the number of terms that are needed
for the symmetrisation is understood.\footnote{Note that this and several other conventions used in the present paper differ from those of \cite{spin3}.} We also show that the action \eqref{action}
-- while manifestly diffeomorphism invariant -- is invariant under the gauge transformations
\begin{subequations} \label{gauge_intro}
\begin{align}
 \d\vf_{\m\n\r} &=\, 3\,\nabla_{\!(\m}\, \x_{\,\n\r)} \,+\,
\cO\!\left(\vf^2\right) \, , \label{dphi-x-t}\\[10pt]
\d g_{\m\n} \,& =\, 12 \, \x^{\rho\sigma}\, \Big\{\  \nabla_{\!\rho}\, \vf_{\m\n\sigma} \,-\,2\, \nabla_{\!(\m}\, \vf_{\n)\,\rho\sigma} \,+\,2\, g_{\rho(\m|} \left[\, 
\nabla\cdot\vf_{|\n)\,\sigma} -\, \nabla_{\sigma}\, \vf_{|\n)}-\, \nabla_{\!|\n)}\, \vf_{\sigma}\,\right] \nn \\
& \phantom{=\, 6 \, \x^{\rho\sigma}} +\, \12\, g_{\rho\m}\, g_{\sigma\n}\, \nabla\cdot \vf \,-\, g_{\m\n} \left[\,  \nabla\!\cdot \vf_{\rho\sigma}
-\, 2\, \nabla_{\!\rho}\, \vf_{\sigma}\,\right] \Big\} \,+\, \cO\!\left(\vf^3\right) \, , \label{dg-x-t}
\end{align}
\end{subequations}
generated by a traceless $\x^{\rho\sigma}$. It thus preserves the same amount of gauge symmetry as the sum of the 
linearised Einstein-Hilbert and Fronsdal actions.

Our results give further support to the interpretation of
$SL(N,\mathbb{R}) \times SL(N,\mathbb{R})$ CS theories as higher-spin gauge
theories, but at the present stage the metric-like
action \eqref{action} is certainly more involved than its frame-like counterpart
\eqref{CS}. The simplicity of the frame-like action has various
advantages: for instance, it allowed to compute the asymptotic symmetries of
\eqref{CS} on $AdS_3$ spaces \cite{HR,spin3,GH,Wlambda}. The appearance of
non-linear $\cW$-algebras then led to a duality conjecture between a class of higher-spin theories with
matter couplings \cite{VP} and $\cW_N$ minimal models \cite{GG}; see \cite{minimal-rev} for a review.
Another interesting result was the
identification of solutions of the field equations -- flat connections --
that generalise
the BTZ black hole \cite{BTZ,BHTZ,BH1,BH2,BH3,BH4,BH5}. On the other hand, there
are aspects of higher-spin theories, such as gravitational
interactions, which might be
easier to deal with in a metric-like theory. As an example, in this paper
we use the action \eqref{action} to compute the entropy of higher-spin
black holes using Wald's formula \cite{wald}. In a cylindrical coordinate system $(\r,t,\th)$,
for a non-rotating black hole with a horizon at $\r = \r_h$ we find
\begin{equation} \label{entropy}
S \,=\, \frac{A}{4G} \left. \left\{\, 1 \,-\,
\frac{3}{2}\, \frac{(\vf_{\th\th\th})^2}{(g_{\th\th})^3} \,+\,
\cO\!\left(\vf^4\right) \,\right\} \right|_{\r \,=\, \r_h} \, ,
\end{equation}
where $A$ denotes the length of the horizon.

The paper is organised as follows: to arrive at \eqref{action}, we
consider in sec.\ \ref{sec:coupling} the most general action quadratic
in $\vf$, that contains the minimal coupling to gravity of Fronsdal's
action. In sec.\ \ref{sec:CS} we fix the free parameters by requiring
that its field equations are solved by the class of asymptotically
$AdS_3$ extrema of \eqref{CS}
which we constructed in \cite{spin3}.  In sec.\ \ref{sec:algebra} we
comment on the algebra generated by the gauge transformations \eqref{gauge_intro}. A
generalisation to spin-$s$ fields is relegated to Appendix
\ref{sec:spin}.  In sec.\ \ref{sec:symm} we rederive the result of
sec.\ \ref{sec:CS} by a direct elimination of the spin connections
$\o_\m{}^a$ and $\o_\m{}^{ab}$ from \eqref{CS}, and we discuss the map
between frame- and metric-like gauge transformations. Further details
are presented in Appendix \ref{sec:spin3}.  In sec.\ \ref{sec:wald} we
turn to applications of \eqref{action}: we derive \eqref{entropy} and
we compare our result with the proposal of \cite{BH1,BH2}. We close with a discussion of possible future directions.  Our conventions are
collected in Appendix \ref{sec:conventions}.

\section{Minimal coupling of higher spins to gravity}\label{sec:minimal}

In this section we consider actions that contain
the minimal coupling to gravity of Fronsdal's Lagrangian
\cite{Fronsdal}, but we also allow additional terms quadratic in $\vf$
which are manifestly diffeomorphism invariant and with at most two
derivatives.  The frame-like action \eqref{CS} is indeed
diffeomorphism invariant, its linearisation reduces to Fronsdal's action \cite{spin3} and on shell $\o$ can be rewritten in terms of $e$ and its first derivative. We  show that when $D=3$ all actions of this type are also invariant at the lowest order in $\vf$ under a deformation
of the linearised Fronsdal gauge symmetry and are actually related by
field redefinitions. We eventually select the point in the parameter
space of field redefinitions that corresponds to the action \eqref{CS} through the map \eqref{fields}. 

\subsection{General quadratic coupling for a spin-3 field}\label{sec:coupling}

The free propagation of a spin-3 particle in a Minkowski background of dimension $D\geq4$ can be described by the Fronsdal equations \cite{Fronsdal} 
\begin{equation} \label{fronsdal_flat}
\cF_{\m\n\r} \equiv\, \Box\, \vf_{\m\n\r} \,
-\,3\, \pr^\l\pr_{(\m}\,\vf_{\n\r)\l} \,+\,3\, \pr_{(\m}\pr_{\n}\, \vf_{\r)} \,=\, 0 \, .
\end{equation}
They can be derived from an action which is left invariant by the gauge transformations 
\begin{equation}
\d \vf_{\m\n\r} = 3\,\pr_{(\m\,} \x_{\,\n\r)} \, , \qquad \textrm{with} \quad
\x_\l{}^\l =\,0 \, .
\end{equation}
In $D\geq4$ this guarantees the propagation of the correct 
number\footnote{namely $\frac{1}{3!}(D-3)(D-2)(D+2)$}
of d.o.f., while in $D=3$ it implies that there is no local dynamics
associated to \eqref{fronsdal_flat}. In order to couple this system to
gravity it is natural to try minimal coupling, i.e.\ the
substitutions $\h \to g$ and $\pr \to \nabla$, where $g$ is the
space-time metric and $\nabla$ is the Levi-Civita connection. However,
a consistent coupling must preserve all gauge symmetries of the
linearised theory, while the covariantised Fronsdal
tensor\footnote{This definition assumes a conventional choice for
the ordering of covariant derivatives. In the following we will
consider curvature terms as well, so that there is no lack of
generality in resolving the ambiguity in a convenient way.}
\begin{equation} \label{fronsdal}
\cF_{\m\n\r} =\, \Box\, \vf_{\m\n\r} \,
-\, \frac{3}{2} \left(\, \nabla^\l\nabla_{\!(\m\,} \vf_{\n\r)\l} 
+  \nabla_{\!(\m} \nabla^{\l\,} \vf_{\n\r)\l}\,\right) \,
+\, 3\, \nabla_{\!(\m}\nabla_{\!\n}\, \vf_{\r)}
\end{equation}
transforms under 
\begin{equation}\label{deltaphi}
\d \vf_{\m\n\r} = 3\,\nabla_{\!(\m\,} \x_{\,\n\r)} \, , \qquad \textrm{with} \quad
\x_\l{}^\l =\,0 
\end{equation}
as
\begin{equation} \label{varF}
\begin{split}
\d \cF_{\m\n\r} =\, & -\, 6\, \x^{\lambda\sigma} \nabla_{\!(\m|} R_{\,\lambda|\n\r)\sigma}
\,-\,9\, R_{\,\lambda(\m\n|\sigma} \nabla_{\!|\r)}\, \x^{\lambda\sigma} \,+\,
6\,R_{\,\lambda(\m\n|\sigma} \nabla^\lambda\, \x_{|\r)}{}^\sigma \\[2pt]
& -\,6\, \x^\lambda{}_{(\m|} \nabla_{\!\lambda}\, R_{\,|\n\r)} \,+\,
\frac{3}{2}\, R_{\,\lambda(\m|} \nabla^\lambda\, \x_{|\n\r)} \,-\, 9\, R_{\,\lambda(\m} \nabla_{\!\n}\, \x_{\r)}{}^\lambda \, .
\end{split}
\end{equation}
This equation lies at the heart of the 
Aragone-Deser argument against higher-spin interactions \cite{AD}:
when $D\geq 4$ one cannot cancel the contributions in the Riemann
tensor with a $\x$-dependent gauge transformation of the metric,
and even adding to the action non-minimal terms of
the form $R_{\cdots} \vf_{\cdots} \vf_{\cdots}$ does not improve the
situation. In general one can overcome this problem in two ways;
Fradkin and Vasiliev showed in \cite{FV} that in the presence of a
cosmological constant one can cancel \eqref{varF} by adding
higher-derivative contributions to the action (see also
\cite{minimal,vertices-frame}). This solution preserves the invariance
under diffeomorphisms and eventually led to the Vasiliev equations,
that indeed require a non-zero cosmological constant and are
manifestly diffeomorphism invariant. As an alternative, one can abandon the
minimal coupling and consider also the gravitational field as a
fluctuation around a fixed background; this path allowed to identify
non-trivial interactions even in flat space
\cite{old-LC,Metsaev,MMR,ST,JLT}, but now diffeomorphisms (or a proper
deformation thereof) 
have to be recovered order by order in the metric fluctuation.

When $D=3$ the solution is much simpler: the Weyl tensor vanishes, so
that the dangerous terms are actually proportional to the Ricci tensor
and can be canceled by a $\x$-dependent transformation of the metric
(even in flat space). 
Let us thus consider the action
\begin{equation} \label{Itot}
I \,=\, \frac{1}{16\p G} \int d^3x\, \sqrt{-g} \left(\, \cL_{EH} \,+\, \cL_{F}\,\right) \, ,
\end{equation}
where $\cL_{EH}$ denotes the Einstein-Hilbert Lagrangian,
\begin{equation} \label{LEH}
\cL_{EH} \,=\, R \,+\, \frac{2}{\ell^2} \ ,
\end{equation}
while
\begin{equation} \label{LF}
\begin{split}
\cL_F \,& =\, \vf^{\,\m\n\r} \left(\,
\cF_{\m\n\r} \,-\, \frac{3}{2} \, g_{(\m\n}\, \cF_{\r)}\,\right)
\,+\, \frac{1}{\ell^2}\, \Big(\, m_1\ \vf_{\m\n\r}\, \vf^{\,\m\n\r} 
\,+\, m_2\ \vf_{\m}\, \vf^{\,\m} \,\Big) \\[5pt]
& +\, 3\, R_{\rho\sigma}\, \Big(\, 
k_1\  \vf^{\,\rho}{}_{\m\n}\, \vf^{\,\sigma\,\m\n} 
\,+\, k_2\ \vf^{\,\rho\sigma}{}_{\m}\, \vf^{\,\m} 
\,+\, k_3\ \vf^{\,\rho}\, \vf^{\,\sigma} \, \Big)  \\[5pt]
& +\, 3\, R\, \Big(\, k_4\ \vf_{\m\n\r}\, \vf^{\,\m\n\r} 
\,+\, k_5\ \vf_{\m}\, \vf^{\,\m} \,\Big) \,+\,
\cO\!\left(\vf^4\right) \, .
\end{split}
\end{equation}
This is the most general Lagrangian quadratic in $\vf$ that reduces to the Fronsdal one upon linearisation, is manifestly diffeomorphism invariant and contains at most two derivatives. As already
recalled, the restriction on the number of derivatives is dictated by
our goal to identify the metric-like counterpart of the frame-like
action \eqref{CS}. On dimensional grounds one could also add terms in
$\ell^{-1}$ and a single covariant derivative, but the only candidate is
$\ell^{-1} \nabla\cdot\vf_\l{}^\l$, a total derivative. Furthermore,
higher-order corrections are at least quartic in $\vf$,
as one cannot build a scalar with the inverse metric, three
spin-3 fields and two derivatives. Cubic contributions proportional to
$\ell^{-1}$ and with a single derivative would be available, but they
are not needed for the gauge invariance, and they cannot be generated by 
the elimination of $\o$ from the action \eqref{CS}.

Under covariantised gauge transformations \eqref{deltaphi}, 
the Lagrangian varies as
\begin{equation} \label{var-abstract}
\d \big(\sqrt{-g}\,(\cL_{EH}+\cL_{F})\big) = \left(\frac{\d (\sqrt{-g}\,\cL_{EH})}{\d g_{\m\n}}\, \d g_{\m\n} +
\frac{\d (\sqrt{-g}\,\cL_{F})}{\d \vf_{\m\n\r}}\, \d \vf_{\m\n\r} \right) +
\frac{\d (\sqrt{-g}\,\cL_{F})}{\d g_{\m\n}}\, \d g_{\m\n} \, .
\end{equation}
The terms in parentheses could cancel, up to total
derivatives, if the metric transforms with a gauge
transformation linear in $\vf$ and $\x$. This is possible for any
choice of the coefficients $k_i$, while the ``mass'' coefficients in
\eqref{LF} have to satisfy\footnote{On $(A)dS_3$ backgrounds all
contributions of the form $R_{\cdots}\vf_{\cdots}\vf_{\cdots}$ become 
mass-like terms proportional to $\ell^{-2}$ as the ones that we already
included in \eqref{LF}. Therefore not all parameters in \eqref{LF} can
be free: on $(A)dS_3$ one has to recover the Fronsdal ``mass''
\cite{Fronsdal-AdS}, whose gauge variation cancels~\eqref{varF} on
constant-curvature backgrounds. This is guaranteed by \eqref{masses}.}
\begin{equation} \label{masses}
m_1 \,=\, 6 \left(k_1 + 3 k_4 - 1\right) \, , \qquad m_2 \,=\, 6 \left(k_2 + k_3 + 3 k_5 + \frac{9}{4} \right) \, .
\end{equation}
One can indeed check that when \eqref{masses} holds, the action
\eqref{Itot} is invariant, at lowest order in $\vf$, under \eqref{deltaphi}
and under the simultaneous transformation 
\begin{equation} \label{dg}
\begin{split}
\d g_{\m\n} =\, 3 & \, \bigg\{\, 
2k_2\ \vf_{\m\n\rho} \nabla\cdot \x^{\rho} 
+\, a\,(2k_1+5)\ \vf_{\rho\sigma(\m} \nabla_{\!\n)}\, \x^{\rho\sigma} 
+\, 2\,(2k_1-3)\ \vf_{\rho\sigma(\m} \nabla^\rho\, \x_{\n)}{}^{\sigma} \\
&\! +\, (4k_3+3)\ \vf_{(\m} \nabla\cdot \x_{\n)} 
\,+\, 2b\,(k_2+4)\ \vf_{\rho} \nabla_{\!(\m}\, \x_{\n)}{}^{\rho} 
\,+\, k_2\ \vf_{\rho} \nabla^\rho\, \x_{\m\n} \\[12pt]
&\! +\, 4\ \x^{\rho\sigma}\, \nabla_{\!\rho}\, \vf_{\m\n\sigma} 
\,+\, (a-1)\,(2k_1+5)\ \x^{\rho\sigma}\, \nabla_{\!(\m}\, \vf_{\n)\,\rho\sigma} 
\,+\, 8\ \x^{\rho}{}_{(\m} \nabla\cdot\vf_{\n)\,\rho} \\[7pt]
&\! -\, 8\ \x_{\rho\,(\m} \nabla^{\rho}\, \vf_{\n)}
\,+\, 2(b-1)\,(k_2+4)\ \x^{\rho}{}_{(\m} \nabla_{\!\n)}\, \vf_{\rho} 
\,+\, 2\ \x_{\m\n} \nabla\cdot \vf \\[10pt]
&\! -\, g_{\m\n}\, \Big[\ 3\,(2k_1+4k_4-1)\, \vf_{\rho\sigma\lambda} \nabla^\rho \x^{\sigma\lambda}
\,+\, 4\, \x^{\rho\sigma}\, \nabla\!\cdot \vf_{\rho\sigma} \\
&\! +\, (4k_2 + 4k_3 + 8k_5 + 3)\, \vf_{\rho}\, \nabla\!\cdot \x^{\rho}  
-\, 8\, \x^{\rho\sigma}\, \nabla_{\!\rho}\, \vf_{\sigma} \ \Big] \bigg\}\,.
\end{split}
\end{equation}
The two parameters $a$ and $b$ are undetermined as they parameterise
field dependent diffeomorphisms.

The last term in \eqref{var-abstract} does not vanish and higher-order
corrections to both the
action and the gauge transformations are needed to preserve the gauge symmetry. In secs.~\ref{sec:CS} and \ref{sec:algebra} we shall give further arguments for this.

We found that the coefficients $k_i$ are free, but this does not mean
that we have a five-parameter family of interacting
theories. Actually, one can remove all
$R_{\cdots}\vf_{\cdots}\vf_{\cdots}$ contributions 
with the field redefinition
\begin{equation} \label{redef}
\begin{split}
g^{\text{(new)}}_{\mu\nu} = g_{\m\n} 
\,& -\, 3\, \bigg\{\, k_1\, \vf_{\rho\sigma(\m\,} \vf_{\n)}{}^{\rho\sigma} 
\,+\, k_2\, \vf_{\m\n\rho}\, \vf^{\,\rho}
\,+\, k_3\, \vf_{(\m\,}\vf_{\n)} \\
& -\, g_{\m\n} \left[\, (k_1+2k_4)\, \vf_{\rho\sigma\l}\, \vf^{\rho\sigma\l} 
\,+\, (k_2+k_3+2k_5)\, \vf_{\l}\, \vf^{\,\l} \,\right] \bigg\} \, .
\end{split}
\end{equation}
In conclusion, in three dimensions there exists a two-derivative
coupling between gravity and a spin-3 field, that at
$\mathcal{O}(\vf^2)$ is unique up to field redefinitions. On the other
hand, it is not clear a priori what is the counterpart of the
frame-like action \eqref{CS} under the map \eqref{fields}. This issue
will now be addressed. It will lead to definite values for the $k_i$.

\subsection{Relation with the Chern-Simons action}\label{sec:CS}

It is rather straightforward to construct solutions of the theory in
its CS formulation: the extrema of the action are flat connections $e$
and $\omega$, with the condition that the gravity dreibein is
invertible. The map \eqref{fields} then allows to construct the metric and the spin-3 field explicitly. For a class of asymptotically AdS
solutions this was done in \cite{spin3} leading to
\begin{subequations} \label{solutions}
\begin{align}
g \, & = \ell^2 \frac{d r^2}{r^2} \, - \left\{\, r^2 \, + \, \left(8\pi G  \ell\right)^2\left(\frac{\cL(x^+)\widetilde{\cL}(x^-)}{r^2}
 - \frac{\ell^2}{4}\, \frac{\cW(x^+)\widetilde{\cW}(x^-)}{r^4} \right) \right\} dx^+ dx^- \nn \\
& - \, 8\pi G \ell \left(\, \cL(x^+) (dx^+)^2 \, + \, \widetilde{\cL}(x^-) (dx^-)^2 \,\right) \, , \label{g} \\[10pt]
\vf \, & = -\, \frac{\ell}{8} \,  (8\pi G \ell) \, \Bigg\{ \left( \cW(x^+) (dx^+)^3 \, +
\, \widetilde{\cW}(x^-) (dx^-)^3 \right) \nn \\
& + \, \left( 8\pi G \ell \right) \, \left(\, 2 \,
\frac{\widetilde{\cL}(x^-)\cW(x^+)}{r^2} +  \left( 8\pi G \ell\right)\,
\frac{\cL(x^+)^2\widetilde{\cW}(x^-)}{r^4} \,\right) (dx^+)^2 dx^- \nn \\
& + \, \left( 8\pi G \ell\right) \,\left(\, 2 \, 
\frac{\cL(x^+)\widetilde{\cW}(x^-)}{r^2} +  \left( 8\pi G \ell\right)\,
\frac{\widetilde{\cL}(x^-)^2\cW(x^+)}{r^4} \,\right) (dx^-)^2 dx^+ \Bigg\}\, .
\label{vf}
\end{align}
\end{subequations}
Here $r$ denotes a radial coordinate and 
$x^\pm = \frac{t}{\ell} \pm\th$.

We will now fix the coefficients $k_i$ by
requiring that these fields also solve the equations of motion derived
from the action \eqref{Itot}, to the lowest non-trivial order in $\vf$. This will lead to \eqref{action}.
The equation of motion for the metric is 
\begin{equation} \label{geq}
R_{\m\n} - \frac{1}{2}\, g_{\m\n} R - \frac{1}{\ell^2}\, g_{\m\n}
\,=\, -\, \frac{1}{\sqrt{-g}}\frac{\delta(\sqrt{-g}\,\cL_F)}{\delta g^{\mu\nu}} \, .
\end{equation}
The r.h.s.\ is the energy-momentum tensor of the spin-3 field, whose explicit expression 
is somewhat lengthy and will not be displayed. 
The equation of motion for $\varphi$ is 
\begin{equation} \label{phieq}
\begin{split}
& \cF_{\m\n\r} - \frac{3}{2} \, g_{(\m\n\,} \cF_{\r)} + \left(\,  \frac{m_1}{\ell^2} 
+ 3k_4\, R \,\right) \vf_{\m\n\r} + \left(\, \frac{m_2}{\ell^2} 
+ 3k_5\,R \,\right) g_{(\m\n\,} \vf_{\r)} \\[2pt]
& + 3k_1\, R_{\lambda(\m\,} \vf_{\n\r)}{}^\lambda + \frac{3}{2}k_2\, R_{(\m\n\,}\vf_{\r)}
+ \frac{3}{2}k_2\, g_{(\m\n\,} \vf_{\r)}{}^{\lambda\sigma} R_{\lambda\sigma} 
+ 3k_3\, g_{(\m\n} R_{\r)\sigma}\, \vf^\sigma \,=\, 0 \, .
\end{split}
\end{equation}
In this way we find a unique solution for the coefficients $k_i$:
\begin{equation} \label{numbers}
k_1 \,=\, \frac{3}{2} \, , \qquad k_2 \,=\, 0 \, , \qquad k_3 \,
=\, -\, \frac{3}{4} \, , \qquad k_4 \,=\, -\, \frac{1}{2} \, , \qquad k_5 \,=\, 0 \, .
\end{equation}
We remark that substituting \eqref{numbers} in the gauge transformation of the metric \eqref{dg}, all terms where the derivative acts on the
parameter $\xi^{\r\s}$ vanish. Moreover, the fields \eqref{solutions} solve the equations of motion \eqref{geq} and \eqref{phieq} only at the lowest-order in $\cW$ and $\widetilde{\cW}$. Therefore, one has to add $\cO(\vf^4)$ corrections to the Lagrangian \eqref{LF}.

The solutions \eqref{solutions} are thus
rich enough to fix all coefficients $k_i$, even if they do not
parameterise the whole space of solutions of~\eqref{CS}. For
instance, they do not include black holes with higher-spin charges, as
constructed in \cite{BH1,BH2,BH4,BH5}, but we checked explicitly that
these also solve the metric-like equations of motion. In sec.\
\ref{sec:symm} we confirm the values~\eqref{numbers} for the
coefficients by a direct elimination of the auxiliary fields from the
frame-like action, which guarantees that all fields constructed via
the map \eqref{fields} from solutions of the frame-like theory solve
the equations of motion \eqref{geq}
and \eqref{phieq}. This argument does not depend on the
presence of a cosmological constant, and the coefficients
\eqref{numbers} thus identify the metric-like counterpart of the
action \eqref{CS} also in Minkowski or de Sitter backgrounds.

\subsection{Algebra of gauge transformations}\label{sec:algebra}

We will now present the algebra of metric-like gauge transformations,
up to the accessible orders in $\vf$, and show that it closes on
shell. Recall that the algebra generated by the frame-like
transformations \eqref{gauge_fields} closes off shell, but it still
contains the auxiliary fields which are eliminated, via their
equations of motion, in the metric formulation (see sec.\
\ref{sec:symm}).  To compute the algebra one has to specify the
tensorial nature of fields and gauge parameters. In analogy with
gravity, we assume that gauge fields are symmetric covariant tensors,
while gauge parameters are symmetric contravariant tensors. Moreover,
we impose the trace constraint on the gauge parameter with a projector
built from the metric.

The additional symmetry of the diffeomorphism invariant action \eqref{Itot} 
(with coefficients fixed as in \eqref{numbers} for simplicity) can thus be cast in the form\footnote{$v,w$ are
the vector fields which generate diffeomorphisms and $\xi,\kappa$ the tensor fields  
which generate spin-3 transformations.} 
\begin{subequations} \label{gauge}
\begin{align}
\d_\x\, \vf_{\m\n\r} & =\, 3\left(\, g_{\lambda(\m}\,g_{\n|\sigma} \,-\, \frac{1}{3}\ g_{\lambda\sigma}\, g_{(\m\n|} \,\right) \nabla_{|\r)}\, \x^{\lambda\sigma} \,+\, \cO\!\left(\vf^2\right) \, , \label{dphi-x} \\[10pt]
\d_\x\, g_{\m\n} \, & =\, 12 \, \x^{\rho\sigma} \bigg\{\  \nabla_{\!\rho}\, \vf_{\m\n\sigma}\,
-\, 2\,\nabla_{\!(\m}\, \vf_{\n)\,\rho\sigma} 
\,-\, g_{\rho\sigma}\, g^{\lambda\tau} \left[\, \nabla_{\!\lambda}\, 
\vf_{\m\n\tau}\,-\, 2\,\nabla_{\!(\m}\, \vf_{\n)\,\lambda\tau} \,\right] \nn \\[5pt]
& +\, 2\,g_{\rho(\m|}\,g^{\lambda\tau} \left[\, \nabla_{\!\lambda}\, \vf_{|\n)\,\sigma\tau}
-\, \nabla_{\sigma}\, \vf_{|\n)\lambda\tau} 
-\, \nabla_{\!|\n)}\, \vf_{\sigma\lambda\tau} \,\right]
\,-\, g_{\m\n}\, g^{\lambda\tau} \left[\, \nabla_{\!\lambda}\, \vf_{\rho\sigma\tau}\,
-\, 2\, \nabla_{\!\rho}\, \vf_{\sigma\lambda\tau} \,\right] \nn \\[5pt]
& +\, \12 \left(\, g_{\m\rho} g_{\n\sigma} \,-\, g_{\m\n}
g_{\rho\sigma} \,\right) g^{\lambda\gamma} g^{\alpha\beta}\,
\nabla_{\!\lambda}\, \vf_{\alpha\beta\gamma} \bigg\} \,+\, \cO\!\left(\vf^3\right)\, , \label{dg-x}
\end{align}
\end{subequations}
where now $\x^{\rho\sigma}$ is a \emph{traceful} tensor. The extra terms appearing in \eqref{gauge} compared to \eqref{gauge_intro} are those needed to implement the traceless projection, i.e.\ they are introduced by the substitution
\begin{equation} \label{g-projector}
\x^{\m\n} \, \to \, \left( \d_\rho^\m \d_\sigma^\n \,-\, \frac{1}{3}\, g_{\rho\sigma}\,g^{\m\n} \right) \x^{\rho\sigma} \, .
\end{equation}
The issue is that the trace condition involves the metric which itself transforms under a second 
transformation on $\delta_\xi\varphi$.

Diffeomorphisms generate an off-shell subalgebra since they satisfy
\begin{equation} \label{22-comm}
[\, \d_v , \d_w \,]\, g_{\m\n} =\, \d_{[v,w]}\, g_{\m\n} \, ,  \qquad\qquad [\, \d_v , \d_w \,]\, \vf_{\m\n\r} =\, \d_{[v,w]}\, \vf_{\m\n\r} \, ,
\end{equation}
where 
\begin{equation} \label{lie1}
[\, v , w \,]^\m =\, v^\n \pr_\n\, w^\m \,-\, w^\n \pr_\n\, v^\m
\end{equation}
is the Lie bracket of the two vector fields $v$ and $w$. 
The commutator of a diffeomorphism with a spin-3 transformation can be cast in a similar form. For instance,
\begin{equation} \label{23-comp}
[\, \d_v , \d_\x \,]\, \vf_{\m\n\r}  =\, 3\left(\, g_{\lambda(\m}\,g_{\n|\sigma} \,-\, \frac{1}{3}\ g_{\lambda\sigma}\, g_{(\m\n|} \,\right) \nabla_{|\r)}\, [\, v , \x \,]^{\lambda\sigma} \, .
\end{equation}
The resulting spin-3 transformation is generated by the Lie derivative of $\x^{\m\n}$ along $v^\s$,
\begin{equation} \label{lie2}
[\, v , \x \,]^{\mu\nu} =\, v^\s \pr_\s \x^{\mu\nu} \,-\, 2\,\x^{\s(\mu|} \pr_\s v^{|\nu)} \, .
\end{equation}
A similar result holds for the metric, so that
\begin{equation} \label{23-comm}
[\, \d_v , \d_\x \,]\, g_{\m\n} =\, \d_{[v,\x]}\, g_{\m\n} \, ,  \qquad\qquad [\, \d_v , \d_\x \,]\, \vf_{\m\n\r} =\, \d_{[v,\x]}\, \vf_{\m\n\r} \, .
\end{equation}
Note that $[\, v , \x \,]^{\mu\nu}$ is not traceless even for a traceless $\x^{\mu\nu}$, but this is not a problem thanks to the projector which multiplies it. Moreover, \eqref{23-comm} remains true in arbitrary space-time dimensions. The peculiarities of the three-dimensional case thus manifest themselves in the commutator of two spin-3 transformations.

We were able to evaluate the commutators \eqref{22-comm} and
\eqref{23-comm} because a diffeomorphism generates terms that are linear in the field on which it acts. Therefore, different orders in the expansion in powers of $\vf$ cannot mix, and we expect that \eqref{23-comm} continue to hold order by order in $\vf$. The situation is very different when one considers commutators of two spin-3 transformations: these are not linear in the fields and, as a result, different orders could mix in the commutator. When the commutator acts on the spin-3 field, the higher-order corrections to \eqref{gauge} could even have an effect at lowest order.
In fact, the gauge variation of $\vf$ is schematically of the form
\begin{equation}
\d \vf \,\sim \, g^2\, \nabla\, \x \,+\, \cO\!\left(\vf^2\right) \, .
\end{equation}
In a second gauge transformation the $\cO\!\left(\vf\right)$ variation of the metric mixes with the corresponding terms coming from the variation of the (yet unknown) $\cO\!\left(\vf^2\right)$ corrections. The only conclusion that we can draw from this commutator is that we do need corrections, since in their absence it is not possible to obtain a diffeomorphism out of 
$[\, \d_\k , \d_\x \,]\, \vf_{\m\n\r}$. On the contrary, when the commutator acts on the metric there is a term of order zero in $\vf$ that is insensitive to any correction to \eqref{gauge} and reads
\begin{equation} \label{step1}
\begin{split}
& [\, \d_\k \, , \d_\x \,]\, g_{\m\n} \, =\, 12\, \x^{\rho\sigma}\, \Big\{\, \nabla_{\!\rho} \nabla_{\!\sigma}\, \k_{\m\n} 
+ 2\,\nabla_{\!\rho} \nabla_{\!(\m} \k_{\n)\sigma} 
- 4\, \nabla_{\!(\m|} \nabla_{\!\rho}\, \k_{|\n)\sigma} 
- 2\,\nabla_{\!(\m} \nabla_{\!\n)} \k_{\rho\sigma} \\[5pt]
& +\,2\, g_{\rho(\m|} \left[\, \Box \k_{|\n)\sigma} 
+ \nabla_{\!\lambda}\nabla_{\!|\n)} \k_\sigma{}^\lambda 
+ \nabla_{\!\lambda}\nabla_{\!\sigma}\, \k_{|\n)}{}^\lambda 
- 2\, \nabla_{\!|\n)}\nabla_{\!\lambda}\, \k_\sigma{}^\lambda 
- 2\, \nabla_{\!\sigma}\nabla_{\!\lambda}\, \k_{|\n)}{}^\lambda  \,\right] \\[5pt]
& -\, g_{\m\n} \left[\, \Box \k_{\rho\sigma} 
+ 2\, \nabla_{\!\lambda}\nabla_{\!\rho}\, \k_\sigma{}^\lambda 
- 4\, \nabla_{\!\rho}\nabla_{\!\lambda}\, \k_\sigma{}^\lambda \,\right] \,
+\, g_{\m\rho} g_{\n\sigma} \nabla_{\!\lambda}\nabla_{\!\tau}\, \k^{\lambda\tau} \,\Big\} \\[4pt]
& -\, \left(\, \x \leftrightarrow \k \,\right)\,+\, \cO\!\left(\vf^2\right) \, .
\end{split}
\end{equation}
For simplicity we presented the result in terms of traceless parameters; inserting \eqref{g-projector} one recovers the full expression.
At this order the right-hand side cannot contain a spin-3 transformation (that is at least linear in $\vf$), so that it must be a diffeomorphism in order to grant the closure of the algebra. This can be made manifest rewriting \eqref{step1} as
\begin{equation} \label{final}
\begin{split}
& [\, \d_\k \, , \d_\x \,]\, g_{\m\n} \,=\, -\, 36\, \Big\{\, \nabla_{\!(\m|} \left(\, \x^{\rho\sigma} \nabla_{\!|\n)}\, \k_{\rho\sigma} - \k^{\rho\sigma} \nabla_{\!|\n)}\, \x_{\rho\sigma} \,\right) \\[5pt]
& -\, 3 \left(\, \x^{\rho\sigma} R_{\rho\sigma}\, \k_{\m\n} - \k^{\rho\sigma} R_{\rho\sigma}\, \x_{\m\n} \,\right) +\, 4 \left(\, \x^{\rho\sigma} R_{\rho(\m\,} \k_{\n)\sigma} - \k^{\rho\sigma} R_{\rho(\m\,} \x_{\n)\sigma} \,\right) \\[5pt]
& -\, \x^{\rho\sigma}\, Y^{\,T}_{\{2,2\}} \nabla_{\!(\rho}\nabla_{\sigma)} \k_{\m\n} + \k^{\rho\sigma}\, Y^{\,T}_{\{2,2\}} \nabla_{\!(\rho}\nabla_{\sigma)} \x_{\m\n} \,\Big\} \,+\, \cO\!\left(\vf^2\right) \, ,
\end{split}
\end{equation}
where $Y^{\,T}_{\{2,2\}}$ denotes the projector onto the traceless $\{2,2\}$ irrep of the group of permutations acting on the space-time indices $\m,\n,\rho,\sigma$. It can be built in terms of
\be
\begin{split}
\cT_{\m\n,\,\r\s} \equiv\, Y_{\{2,2\}} \nabla_{\!(\m} \nabla_{\!\n)\,} \x_{\r\s} = \frac{1}{6}\, \big(\, 2\, & \nabla_{\!(\m} \nabla_{\!\n)\,} \x_{\r\s} - \nabla_{\!\r} \nabla_{\!(\m\,} \x_{\n)\s} - \nabla_{\!\s} \nabla_{\!(\m\,} \x_{\n)\r}  \\
 - \,& \nabla_{\!\m} \nabla_{\!(\r\,} \x_{\s)\n} - \nabla_{\!\n} \nabla_{\!(\r\,} \x_{\s)\m} + 2\, \nabla_{\!(\r} \nabla_{\!\s)\,} \x_{\m\n} \,\big)
\end{split}
\ee
as
\begin{equation}
\begin{split}
Y^{\,T}_{\{2,2\}} \nabla_{\!(\m} \nabla_{\!\n)\,} \x_{\r\s} =\, \cT_{\m\n,\,\r\s} \, & - \left(\, g_{\m\n} \cT_{\r\s} - g_{\r(\m} \cT_{\n)\s} - g_{\s(\m} \cT_{\n)\r} + g_{\r\s} \cT_{\m\n} \,\right)\\
& + \frac{1}{2} \left(\, g_{\m\n}g_{\r\s} - g_{\r(\m}g_{\n)\s} \,\right) \cT \, ,
\end{split}
\end{equation}
where we defined $\cT_{\m\n} \equiv \cT_{\m\n,\, \l}{}^\l$ and $\cT \equiv \cT_{\r}{}^{\r}{}_{,\, \s}{}^\s$. The resulting combination has the same symmetries as the Weyl tensor and thus vanishes in three dimensions. In conclusion, we recovered a diffeomorphism generated by\footnote{With respect to \eqref{step1} and \eqref{final} we reinserted here the full dependence on the metric, so that $\k^{\a\b}$ and $\x^{\a\b}$ are traceful tensors.}
\begin{equation}
v^\m \,=\, 18\, g^{\m\nu} \left\{\, \k^{\rho\sigma}\nabla_{\!\nu\,} \x_{\rho\sigma} \,-\, \x^{\rho\sigma}\nabla_{\!\nu\,} \k_{\rho\sigma} \,-\, \frac{1}{3} \left(\, \k^\rho{}_\rho \nabla_{\!\nu\,} \x^\sigma{}_\sigma \,-\, \x^\rho{}_\rho \nabla_{\!\nu\,} \k^\sigma{}_\sigma \,\right)\,\right\} \, ,
\end{equation}
plus a remainder proportional to the Ricci tensor. However, one can rewrite it in terms of the Einstein tensor: if one performs the substitution
\begin{equation}
R_{\mu\nu} \to R_{\mu\nu} \,-\, \12\ g_{\mu\nu} \left( R \,+\, \frac{2}{\ell^2} \right)
\end{equation}
the contributions in $g_{\mu\nu}$ cancel either identically or on account of the tracelessness of the parameters. At this order in $\vf$ the field equations for the metric are the Einstein equations, so that the remainder in \eqref{final} signals that the algebra of gauge transformations closes only on shell.

\section{Mapping frame- to metric-like formulation}\label{sec:symm}

In the last section we identified the metric-like theory that
corresponds to the higher-spin theory defined by the frame-like
action~\eqref{CS} to the first non-trivial order by analysing its
equations of motion. Alternatively, one can identify it directly by eliminating
the spin connections. This will be the content of this section. This
also leads to an identification between the parameters
of frame- and metric-like gauge symmetries.

\subsection{Elimination of spin connections}\label{sec:elimination}

To go from the frame-like formulation to the metric-like formulation,
we have to solve the torsion constraint (the equation of motion for
$\omega$ derived from the action~\eqref{CS}),
\begin{equation}
de + \omega \wedge e +e\wedge \omega = 0 \ ,
\end{equation}
to express the generalised spin connection $\omega$ in terms of the 
generalised vielbein $e$. In components this constraint reads
\begin{equation}\label{torsionconstraint}
\partial^{\phantom{\cC}}_{[\mu}\, e_{\nu ]}^{\cC} + f_{\cA \cB}{}^{\cC}\, 
\omega_{[\mu}^{\cA}\, e_{\nu]}^{\cB} = 0 \qquad \Leftrightarrow \qquad 
\cD_{[\mu}^{\phantom{\cC}}e^{\cC}_{\nu]} = 0 \ ,
\end{equation}
where the curly letters are labels for the generators $\{J_{\cA}\}$ of
$sl(3,\mathbb{R})$, and $\cD_{\mu}$ denotes the full $sl(3)$ covariant connection
including Levi-Civita and $sl(3)$ spin connection. The
torsion constraint \eqref{torsionconstraint} together with the metric
being covariantly constant,
\begin{equation}
\nabla_{\!\mu\,} g_{\nu\rho} \, = \, \cD_{\mu\,} g_{\nu\rho} \, = \, 0 \ ,
\end{equation}
lead to 
\begin{equation}\label{genvielbeinpostulate}
\kappa_{\cA\cB} \, e^{\cA}_{\rho}\, \cD_{\mu\,}^{\phantom{\cB}}e^{\cB}_{\nu} \,= \,0 \ .
\end{equation}
For $\mathfrak{g}=sl(2,\mathbb{R})$ one could multiply this expression by the
inverse vielbein to conclude that the vielbein is covariantly constant
under the full connection (vielbein postulate), which allows to solve
for the spin connection. For general Lie algebras this is not directly
possible, and instead we will solve for the spin connection in a
perturbative expansion. For that we separate the generators into the
$sl(2)$ generators $\{J_{a}\}$ (labelled by small Latin indices), and
the remaining generators $\{J_{A}\}$ (labelled by capital Latin
indices and chosen to be orthogonal to the $J_{a}$ with respect to the
Killing form). The relation~\eqref{genvielbeinpostulate} can be separated,
and one obtains
\begin{equation}\label{corrvielbeinpostulate}
\cD_{\mu}^{\phantom{c}}\,e_{\nu}^{c} = -\, \kappa_{AB}\,\os{0}{g}{}^{\rho\lambda}\,
e^{c}_{\lambda}\,e^{A}_{\rho}\,\cD_{\mu}^{\phantom{B}}e_{\nu}^{B} \ ,
\end{equation}
where we used the inverse $\os{0}{g}{}^{\mu \nu}$ of the
$sl(2)$-part $\os{0}{g}{}_{\mu \nu}$ of the metric, 
\begin{equation}
\os{0}{g}{}_{\mu \nu} \equiv  e_{\mu}^{a}\, e_{\nu}^{b}\, \kappa_{ab} \ .
\end{equation}
Eq.\ \eqref{corrvielbeinpostulate} can be interpreted as the
correction to the vielbein postulate of the $sl(2)$ components $e^{c}$
of the vielbein. It remains to also get an expression for the covariant derivative of the
non-$sl(2)$ components $e^{C}$, and for that we have to determine the spin
connection. To this end we split the sum over $\cB = (b,B)$
in~\eqref{torsionconstraint} and we arrive, after some simple
manipulations, at
\begin{equation}\label{torsionconstraint2}
f_{mn}{}^{p}\,f_{\cA}{}^{\cC m}\, \os{0}{g}{}^{\mu \nu}\,
\omega^{\cA}_{\mu}\, e^{n}_{\nu} 
= -\, f_{mn}{}^{p}\,
\os{0}{g}{}^{\mu\rho}\, \os{0}{g}{}^{\nu \sigma}\, e^{m}_{\rho}\,
e^{n}_{\sigma}\, \left( \partial_{[\mu}^{\phantom{\cC}}\, e_{\nu]}^{\cC} + f_{\cA
B}{}^{\cC}\, \omega_{[\mu}^{\cA}\, e_{\nu]}^{B} \right) \equiv V^{\cC,p} \ .
\end{equation}
We will now solve~\eqref{torsionconstraint2} for $\omega$ in terms of
$V$. By replacing the result for $\omega$ in the $\omega$-dependence
of $V$ successively, we arrive at a perturbative expansion in the number
of higher-spin vielbeins $e^{B}$.

We contract~\eqref{torsionconstraint2} with a suitable combination of
structure constants, and arrive, after repeated use of the Jacobi identity and 
other identities collected in Appendix \ref{sec:conventions}, at
\begin{equation}
(C_{2})^{\cD}_{\cA}\,\omega^{\cA}_{\mu}\, e^{b}_{\nu}\, \os{0}{g}{}^{\mu\nu}
= \left(\delta_{a}^{b}\,\delta_{\cC}^{\cD}
+f_{a}{}^{be}\, f_{e\cC}{}^{\cD}
+f_{a \cC}{}^{\cE}\, f_{\cE}{}^{b\cD}
\right) V^{\cC,a}\ .
\end{equation}
Here, $C_{2}$ denotes the quadratic Casimir of $sl(2)$ in the adjoint
representation on $sl(3)$,
\begin{equation}
(C_{2})_{\cA}^{\cD} = f^{\cD}{}_{m\cC}\, f^{\cC m}{}_{\cA} \ .
\end{equation}
The Casimir can of course be inverted as it is constant on
every irreducible representation that occurs.
In our convention the Casimir takes the value $s(s-1)$ if the
corresponding $sl(2)$-representation has dimension $(2s-1)$. Note that
until now all manipulations were valid for general semi-simple Lie algebras. We
now specialise to $sl(3)$, and we find
\begin{align}
2\,\omega^{d}_{\beta}
&= e^{b}_{\beta}\, \left(\kappa_{ab}\,\delta_{c}^{d}
+f_{ab}{}^{e}\, f_{ec}{}^{d}
+f_{a c}{}^{e}\, f_{eb}{}^{d} \right) V^{c,a} \ , \\[5pt]
6\,\omega^{D}_{\beta} 
&= e^{b}_{\beta}\, \left(\kappa_{ab}\,\delta_{C}^{D}
+f_{ab}{}^{e}\, f_{eC}{}^{D}
+f_{a C}{}^{E}\, f_{Eb}{}^{D}
\right) V^{C,a}\ .
\end{align}
In the following, we will use the notation $e_{\mu}^{a}$ for $sl(2)$
components of the vielbein, and $E_{\mu}^{A}$ for the remaining
components to better distinguish them. Similarly we use
$\omega^{a}_{\mu}$ and $\Omega^{A}_{\mu}$. The lowest-order solutions
for the spin connections then read
\begin{subequations}\label{spinconnections}
\begin{align}
\omega^{d}_{\beta}
&= -\frac{1}{2}e^{b}_{\beta}\, 
\left(f_{mnb}\, \delta_{c}^{d} 
+2\,\kappa_{b[n}\, f_{m]c}{}^{d}
+2\,\kappa_{c[n}\, f_{m]b}{}^{d} \right)
\os{0}{g}{}^{\mu\rho}\,
\os{0}{g}{}^{\nu \sigma}\, e^{m}_{\rho}\, e^{n}_{\sigma}\, 
\partial_{[\mu}^{\phantom{c}}\, e_{\nu]}^{c} +\mathcal{O}\!\left(E^{2}\right) \, , \\
\Omega^{D}_{\beta}
&=  -\frac{1}{6}e^{b}_{\beta}\, f_{mn}{}^{a}\,\left(\kappa_{ab}\,\delta_{C}^{D}
+f_{ab}{}^{e}\, f_{eC}{}^{D}
+f_{a C}{}^{E}\, f_{Eb}{}^{D}
\right)\nonumber\\
&\qquad\qquad \times 
 \os{0}{g}{}^{\mu\rho}\,
\os{0}{g}{}^{\nu \sigma}\, e^{m}_{\rho}\, e^{n}_{\sigma}\, 
\left( \partial_{[\mu}^{\phantom{C}}\, E_{\nu]}^{C} 
+ f_{a B}{}^{C}\, 
\omega_{[\mu}^{a}\, 
E_{\nu]}^{B} \right) \ .
\label{spin3connection}
\end{align}
\end{subequations}
Note that the second equation does not receive corrections of higher
order in $E$,\footnote{There are no
$\Omega^{A}$ appearing on the right-hand side due
to the fact that in $sl(3)$ the structure constants involving only
non-$sl(2)$ indices vanish, $f_{ABC}=0$.} but replacing
$\omega^{a}$ by its lowest order contribution will only yield
$\Omega^{D}$ up to terms of order $\mathcal{O} (E^{3})$.

We could use these relations to rewrite the frame-like
action~\eqref{CS} in terms of vielbeins, and then rewrite it in terms
of metric $g$ and spin-3 field $\varphi$. Instead we will follow a
different route by determining the gauge transformations of the
metric-like theory from those in the frame-like theory.

\subsection{Relations between gauge parameters}

The frame-like theory is invariant under the gauge
transformations~\eqref{gauge_fields}, i.e.\ under generalised Lorentz
transformations,
\begin{equation}
\delta e_{\mu}^{\cC}\, =\, f_{\cA \cB}{}^{\cC}\,
e^{\cA}_{\mu}\, \Lambda^{\cB}\ ,
\end{equation}
and generalised local translations,
\begin{equation} \label{trasl}
\d e_\m^{\cC} \,=\, \pr_\m \, \x^{\cC} \,+\, f_{\cA\cB}{}^{\cC }\,
\o_\m^{\cA}\, \x^{\cB}  \ .
\end{equation}
In the frame-like description of pure gravity, diffeomorphisms are
induced by local translations (see e.g.\ \cite{witten}). The same
argument applies to the higher-spin setup: a local translation by
\begin{equation} \label{par-diffeo}
\x^{\cA} \,=\, e_\m^{\cA}\, \xi^\m \, ,
\end{equation}
can be decomposed as 
\begin{equation} 
\d e_\m^{\cC} = \, e_{\lambda}^{\cC}\, \partial_{\mu}\xi^{\lambda} +
\xi^{\lambda}\,\partial_{\lambda}e_{\mu}^{\cC}
+2\,\xi^{\lambda} \left(\partial_{[\mu}^{\phantom{\cC}}e_{\lambda ]}^{\cC} 
+ f_{\cA\cB}{}^{\cC}\, \omega^{\cA}_{[\mu}\,e^{\cB}_{\lambda]} 
\right) 
 +\xi^{\lambda}\,f_{\cA\cB}{}^{\cC}\,
\omega_{\lambda}^{\cA}\,e^{\cB}_{\mu} \ .
\end{equation}
The term in parentheses is proportional to the torsion
constraint~\eqref{torsionconstraint}, so that up to a Lorentz-like
transformation generated by
$\Lambda^{\cB}=-\,\xi^{\lambda}\omega^{\cB}_{\lambda}$ the transformation is on-shell
equivalent to a diffeomorphism generated by $-\xi^{\mu}$. 

In the pure gravity case, one can invert this argument to conclude
that any local translation by $\xi^{a}$ generates a diffeomorphism
where the corresponding vector field is obtained by contracting
$\xi^{a}$ with the inverse vielbein. In the higher-spin theory we expect
that this expression is modified, and the simplest ansatz covariant
in the $sl(3)$ indices leads to
\begin{equation}\label{diffeofromframe}
\xi^{\mu} = \, g^{\mu\nu}\, \kappa_{\cA\cB}\, e_{\nu}^{\cA}\, \xi^{\cB} \ .
\end{equation}
This is consistent with the previous argument: if we insert
$\xi^{\cA}$ given in~\eqref{par-diffeo} into~\eqref{diffeofromframe}
we obtain an identity. 

A general gauge parameter $\xi^{\cA}$ will induce a spin-3
transformation as well as a diffeomorphism given
by~\eqref{diffeofromframe}. On the other hand, we know
from~\eqref{par-diffeo} that the diffeomorphism part is generated by
\begin{equation}
\tilde{\xi}^{\cA}=\, e^{\cA}_{\mu}\, 
g^{\mu\nu}\, \kappa_{\cB\cC}\, e_{\nu}^{\cC}\, \xi^{\cB} \equiv
\mathcal{P}^{\cA}_{\ \cB} \, \xi^{\cB}\ ,
\end{equation}
and we can identify $\mathcal{P}^{\cA}_{\ \cB}$ as the projector onto
diffeomorphisms (that it is indeed a projector can be easily
verified). This means in turn that the projector on pure spin-3
transformations is $(1-\mathcal{P})$.

To find the correct ansatz for the spin-3 transformation we start from
the linearised relation for a free spin-3 field,
\begin{equation}
\xi^{ab}=3\, \bar{e}^{a}_{\mu}\, \bar{e}^{b}_{\nu}\, \xi^{\mu\nu}  \ ,
\end{equation}
where $\bar{e}$ denotes the vielbein of the fixed background
(for which $\bar e_\mu^A=0$). We can
rewrite this with the help of the symmetric structure constants
$d_{\cA\cB\cC}$ (see appendix~\ref{sec:conventions}) as
\begin{equation}
\xi^{A}=\, 3\,d^{A}{}_{ab}\,\bar{e}^{a}_{\mu}\, \bar{e}^{b}_{\nu}\,
\xi^{\mu\nu}  \ .
\end{equation}
To obtain the corrections to this expression in the
non-linear theory, we let the indices run over all $sl(3)$ labels and
project the result by $(1-\mathcal{P})$ to ensure that no
diffeomorphism is generated. This leads to the following ansatz for a
pure spin-3 transformation generated by $\xi^{\mu\nu}$,
\begin{align}
\xi^{\cA} &= 3 \left( \delta^{\cA}_{\cB}-\mathcal{P}^{\cA}_{\ \cB}\right) 
\xi^{\mu\nu} \,e^{\cC}_{\mu}\,
e^{\cD}_{\nu} \, d^{\cB}{}_{\cC\cD} \nn \\[5pt]
& = 3\left(\xi^{\mu\nu} \,e^{\cC}_{\mu}\,
e^{\cD}_{\nu} \, d^{\cA}{}_{\cC\cD} - d_{\cB\cC\cD}\,e^{\cA}_{\rho}\,
e^{\cB}_{\sigma}\, e^{\cC}_{\mu}\, e^{\cD}_{\nu}\, 
g^{\rho\sigma}\, \xi^{\mu \nu} \right) \ .
\label{spin3frame}
\end{align}
With this ansatz we can then derive the expression for the spin-3
transformation in the metric-like theory, which will be done in the
following section. Note that this ansatz most likely needs to be
modified at higher order in $E$, which becomes important if one
computes higher order corrections to the gauge transformations.

\subsection{Transformations of metric-like fields}

Coming from the frame-like theory we can derive the spin-3 gauge
transformations in the metric-like theory. The strategy is the
following: we first express the metric
and the spin-3 field in terms of vielbeins (see eq.\ \eqref{fields}),
\begin{equation}\label{fieldsexplicit}
g_{\mu\nu}=\, \kappa_{\cA\cB}\,e^{\cA}_{\mu}\,e^{\cB}_{\nu} \, ,
\qquad \vf_{\mu\nu\rho} =\, \frac{1}{6}\, d_{\cA\cB\cC}\,e^{\cA}_{\mu}\,
e^{\cB}_{\nu}\,e^{\cC}_{\rho} \ ,
\end{equation}
then use the gauge transformations~\eqref{trasl} of the vielbeins
under local translations by a parameter given in~\eqref{spin3frame},
insert the spin connection in terms of the vielbein (see eq.\
\eqref{spinconnections}), and finally express the result in terms of
$g$ and $\vf$.

Let us start with the transformation of the spin-3 field $\vf$. Under
a local translation it transforms as
\begin{equation}
\delta \vf_{\mu \nu \rho} =\, \frac{1}{2}\, d_{\cA\cB\cC}\, 
e^{\cA}_{(\mu}\,e^{\cB}_{\nu}\,\cD_{\rho)}\xi^{\cC} \ .
\end{equation}
Replacing the gauge parameter by the expression~\eqref{spin3frame} and
expanding in powers of $E$ we obtain
\begin{equation}
\delta \vf_{\mu \nu \rho} =\, \frac{3}{2}\, d_{Abc}\, d^{A}{}_{de} \,
e^{b}_{(\mu}\,e^{c}_{\nu}\,\cD_{\rho)}
\left(e^{d}_{\lambda}\,e^{e}_{\sigma}\,\xi^{\lambda\sigma}\right)
+ \mathcal{O} (E^{2}) \ .
\end{equation}
The covariant derivative of the $sl(2)$ components of the vielbein is
of order $\mathcal{O} (E^{2})$ (see eq.\
\eqref{corrvielbeinpostulate}), and we arrive at
\begin{align}
\delta  \vf_{\mu \nu \rho} & =\, \frac{3}{2}\, d_{Abc}\,d^{A}{}_{de}\,
e^{d}_{\lambda}\,e^{e}_{\sigma}\,
e^{b}_{(\mu}\,e^{c}_{\nu}\,\nabla_{\rho)}\xi^{\lambda\sigma} + \mathcal{O}\! \left(E^{2}\right)\nn \\
&=\,3\, \nabla_{(\mu}\left(\xi_{\nu\rho)} -\frac{1}{3}g_{\nu\rho)}\, 
g^{\lambda\sigma}\,\xi_{\lambda\sigma}  \right)+ \mathcal{O} \!\left(E^{2}\right) \ ,
\end{align}
where we used the identity~\eqref{app:dd-identity} for the structure
constants. This result equals~\eqref{dphi-x} including the projection
of $\xi_{\lambda\sigma}$ to its traceless part.

We now consider the spin-3 transformation of the metric. Under a
general local translation~\eqref{trasl} it transforms as 
\begin{equation}
\delta g_{\mu\nu} =\, 2\,\kappa_{\cA\cB}\, e^{\cA}_{(\mu} \,
\cD_{\nu)}\xi^{\cB}\ .
\end{equation}
We replace the gauge parameter by the expression~\eqref{spin3frame} for a pure spin-3
transformation, and after some manipulations
where we also use~\eqref{genvielbeinpostulate} we arrive at
\begin{align}
\delta g_{\mu\nu} &=\, -6\, d_{\cA\cB\cC}\, \xi^{\rho\sigma}\, e^{\cB}_{\rho}\,
e^{\cC}_{\sigma}\,\cD_{(\mu}e^{\cA}_{\nu)}\nn \\[5pt]
\label{varmetricfromframe}
 & =\, -6\, d_{Abc}\, \xi^{\rho\sigma}\,e^{b}_{\rho}\, e^{c}_{\sigma}\,
\mathcal{D}_{(\mu}E_{\nu)}^{A} + \mathcal{O}
\!\left(E^{3}\right) \ ,
\end{align}
where in the last step we expanded in $sl(2)$ and non-$sl(2)$ indices
using that the covariant derivative of $e^{a}$ is of order
$\mathcal{O} (E^{2})$ (see eq.\ \eqref{corrvielbeinpostulate}).  One
observes that in~\eqref{varmetricfromframe} no derivatives of the
gauge parameter $\xi^{\rho\sigma}$ appear. In fact one can show
(see appendix~\ref{sec:spin3} for the details) that this result
precisely reproduces the gauge transformation~\eqref{dg-x}. Since
there is a unique action at quadratic order in $\vf$ that is invariant
under this gauge transformation, we have again identified the
action~\eqref{action} as the metric-like counterpart of the frame-like
action.

\section{Wald entropy for higher-spin black holes}\label{sec:wald}

In ref.\ \cite{BH1} the class of solutions considered in 
\cite{spin3} was enlarged, in search for black holes with
higher-spin charges; see also \cite{BH2,BH3,BH4,BH5}. Although the presence of an event horizon is not a gauge-invariant statement in the theories we are considering (due to the Fronsdal-like transformations \eqref{gauge}), there is a gauge where these solutions exhibit a regular event horizon \cite{BH2}. This gauge is also supposed to be unique, and one can thus try to evaluate the entropy of higher-spin black holes using Wald's formula \cite{wald}. For a static
black hole in three dimensions with metric
\begin{equation} \label{metric-diag}
g \,=\, g_{\r\r}(\r)\, d\r^2 \,+\, g_{tt}(\r)\, dt^2 \,+\, g_{\th\th}(\r)\, d\th^2 
\end{equation}
and regular  horizon at $\r = \r_h$ it reads
\begin{equation} \label{wald2}
S \,=\, \frac{\p}{G} \left. \sqrt{g_{\th\th}}\, g_{tt}\, g_{\r\r}\, \frac{\d \cL}{\d R_{\,t\r\,t\r}} \right|_{\r \,=\, \r_h} .
\end{equation}
In all static solutions considered in \cite{BH2,BH3,BH4,BH5} the spin-3 field takes the form
\begin{equation} \label{spin3-diag}
\vf \,=\, 3\, \vf_{\r\r\th}(\r)\, d\r^2 d\th \,+\,
3\,\vf_{tt\th}(\r)\, dt^2 d\th \,+\, \vf_{\th\th\th}(\r)\, d\th^3 \, .
\end{equation}
Evaluating \eqref{wald2} taking into account \eqref{metric-diag} and \eqref{spin3-diag} we find
\begin{equation}\label{Wald3}
\begin{split}
S \, =&\,  \frac{\p}{2G} \sqrt{g_{\th\th}} \,\bigg\{\, 1 \,
+\,\frac{3}{2} \left[\, 2\,(k_4 + k_5) \big(g^{\th\th}\big)^3\big(\vf_{\th\th\th}\big)^2 \,+\, 2\, (k_2 + 2k_5)\,
g^{tt}g^{\r\r}g^{\th\th} \vf_{tt\th}\,\vf_{\r\r\th} \right. \\
& \qquad\quad  +\, (2k_1 + k_2 + 6 k_4 + 2k_5) \left(  \big(g^{tt}\big)^2  g^{\th\th} \big(\vf_{tt\th}\big)^2 
+  \big(g^{\r\r} \big)^2 g^{\th\th} \big(\vf_{\r\r\th}\big)^2 \right) \\
& \qquad\quad  + \left. (k_2 + 4 k_5)\, \big(g^{\th\th}\big)^2 \left(\, g^{tt}\,  \vf_{tt\th}\, \vf_{\th\th\th} 
+ g^{\r\r}\, \vf_{\r\r\th}\,  \vf_{\th\th\th} \,\right) \,\right] +\, \cO\!\left(\vf^4\right)
\bigg\} \bigg|_{\rho \,=\,\rho_{h}}\, .
\end{split}
\end{equation}
Substituting the values \eqref{numbers} for the coefficients $k_i$ it simplifies
to\footnote{The action \eqref{Itot} can be rewritten in many ways,
e.g.\ by changing the order of covariant derivatives in the Fronsdal tensor which will shift some of the coefficients $k_i$. We have checked that the final
expression for the entropy is unchanged.}
\begin{equation} \label{wald_final}
S \,=\, \frac{A}{4G} \left. \left\{\, 1 \,-\, \frac{3}{2}\, \big(g^{\th\th}\big)^3 \big(\vf_{\th\th\th}\big)^2 \,
+\, \cO\!\left(\vf^4\right) \,\right\} \right|_{\r \,=\, \r_h} \, ,
\end{equation}
where $A=2\pi\sqrt{g_{\theta\theta} (\rho_{h})}$ is the length of the horizon.
The same
result can be recovered by taking advantage of the uniqueness of the two-derivative
coupling up to field redefinitions. The field redefinition
\eqref{redef} cancels all terms with the Ricci tensor 
so that for the new action the black hole entropy is just one quarter of the area of the horizon, i.e. 
\begin{equation}
S \,=\, \frac{\p}{2G}\, \sqrt{g^{\text{(new)}}_{\th\th}(\r_h)} \,
=\,\frac{\p}{2G} \left. \sqrt{g_{\th\th}} \left\{\, 1 \,+\,
\frac{1}{2}\, g^{\th\th} \d g_{\th\th} \,
+\,\cO\!\left(\vf^4\right) \,\right\} \right|_{\r \,=\, \r_h} \, .
\end{equation}
Using \eqref{redef} this can be shown to coincide with \eqref{Wald3}. 

The black hole solutions of \cite{BH1,BH2,BH4,BH5} are constructed in such a way 
that they have a pointwise smooth BTZ limit if one switches off the spin-3 charge. We can then 
parameterise their mass, temperature and entropy by their deviation from the
BTZ limits as 
\begin{subequations} \label{termo}
\begin{align}
M&\,=\,M_{\text{BTZ}}\left(\,1+\a_M\,\e\,\right) \, , \label{M} \\[5pt]
T&\,=\,T_{\text{BTZ}}\left(\,1+\a_T\,\e\,\right) \, , \label{T} \\[5pt]
S&\,=\,S_{\text{BTZ}}\left(\,1+\a_S\,\e\,\right) \, , \label{S}
\end{align}
\end{subequations}
where $\e=0$ in the BTZ limit.  In terms of the dimensionless
parameters $\cL$ and $\cW$ used in~\cite{BH2}, one has
\begin{equation}\label{AGKPsol}
M_{\text{BTZ}}\,=\,\frac{4\pi}{\ell} \cL\,,\qquad 
T_{\text{BTZ}}\,=\,\sqrt{\frac{2\cL}{\pi k\,\ell^{2}}}\,,\qquad
S_{\text{BTZ}}\,=\,4\pi\sqrt{2\pi k\cL} \, ,
\end{equation}
where $k=\frac{\ell}{4G}$, and we set\footnote{Our parameter $\e$ is related to the parameter
$\zeta$ used in~\cite{BH2} by $\e = \frac{32\pi}{k}\zeta^{2}$.}
\begin{equation}
\e\,=\,\frac{\cW^{2}}{\cL^{3}}\, .
\end{equation}
There are also higher-order corrections in $\e$, 
but from our action we can only evaluate the deviation from the BTZ entropy (and the other thermodynamical parameters of the 
black hole) to the lowest non-trivial order in the spin-3
charge. Therefore, we restrict the discussion to $\mathcal{O}(\e)$
terms. In addition there is also the spin-3 charge $Q$. It vanishes as
$\cW\to 0$ and we assume it to be of the form
\begin{equation}
Q \,=\, \cW \left(\, 1 + \a_Q\,\e  + \cdots \,\right)\, .
\end{equation}
However, $\a_Q$ does not affect the following discussion, and to the
order we are computing we can identify $Q$ with $\cW$.

If one expresses the entropy as a function of $M$ and $Q$, the first law of 
black hole thermodynamics states
\begin{equation}\label{firstlaw}
\left(\frac{\partial S}{\partial M}\right)_Q =\, \frac{1}{T} \ .
\end{equation}
We can use \eqref{M} and \eqref{AGKPsol} to obtain 
($\cdots$ are higher order terms in $\cW$)
\begin{equation}
\cL \,=\,\frac{\ell}{4\pi}\,M - (4\pi)^2\alpha_M\,\frac{\cW^2}{\ell^{2}M^2} \,+\, \cdots
\end{equation}
Inserting this into the expression for the entropy we find
\begin{equation}
S \,=\, 2\p \sqrt{2kM\ell} \left(\, 1 + 32\p^3(2\a_S-\a_M)
\frac{\cW^2}{\ell^{3}M^3} + \cdots \,\right) \, ,
\end{equation}
from where we compute 
\begin{equation}
\left(\frac{\partial S}{\partial M}\right)_\cW
=\, \p \sqrt{\frac{2k\ell}{M}} \left(\, 1 - 160\p^3(2\a_S-\a_M)
\frac{\cW^2}{\ell^{3}M^3} + \cdots \,\right) \, .
\end{equation}
On the other hand, for the temperature \eqref{T} we find
\begin{equation}
\frac{1}{T} \,=\, \p \sqrt{\frac{2k\ell}{M}} \left(\, 1 -
32\p^3(2\a_T-\a_M) \frac{\cW^2}{\ell^{3}M^3} + \cdots \,\right) \, .
\end{equation}
Comparison gives the linear relation for the coefficients
\begin{equation}\label{relalphas}
5\alpha_S-2\alpha_M = \a_T \, .
\end{equation}
Demanding regularity at the horizon,
\begin{equation}
\frac{1}{T} \,=\, 2\p
\left.\sqrt{\frac{2\,g_{\r\r}}{-\,g''_{tt}}}\,\right|_{\rho\, =\,\rho_{h}}
\,=\, 2\p
\left.\sqrt{\frac{2\,\vf_{\r\r\th}}{-\,\vf''_{tt\th}}}\,\right|_{\rho\,
=\,\rho_{h}} \, ,
\end{equation}
the temperature is fixed in terms of $\cL$ and $\cW$, and for the solution presented in eqs.\ (4.13) and (C.1) of \cite{BH2} this leads to
\begin{equation}\label{alphaT}
\alpha_T\,=\,-\, \frac{15k}{256\pi} \, .
\end{equation}

We can now evaluate Wald's formula \eqref{wald_final} on the black hole solution of \cite{BH2} to fix $\a_S$. The relation \eqref{relalphas} would then fix also $\a_M$. To this end the overall normalisation of the fields is crucial, and we checked that the metric and the spin-3 field presented in \cite{BH2} solve our equations of motion \eqref{geq} and \eqref{phieq} provided that one multiplies their $\vf_{\m\n\r}$ by $1/6$.\footnote{The need for this rescaling can also be inferred from the comparison between our definition for the metric-like fields \eqref{fields} and the corresponding definition in \cite{BH2}.} Eq.~\eqref{wald_final} eventually implies
\begin{equation}
\alpha_S= \frac{9k}{256\pi} \, ,
\end{equation}
and together with \eqref{relalphas} and \eqref{alphaT}
\begin{equation}
\alpha_M= \frac{15k}{128\pi} \, .
\end{equation}

The correction to the BTZ entropy that we found in this way does not
agree with the one presented in sec.\ 4.2 of \cite{BH2} (see also
\cite{BH3,BH-CFT} for a discussion of the proposal of \cite{BH2} from
a CFT perspective). On the other hand, the previous analysis suggests
a possible interpretation of the mismatch: in \cite{BH2} the entropy
was derived from the first law \eqref{firstlaw} under the assumption
$M = M_{\text{BTZ}}$, i.e.\ $\a_M = 0$. Inserting this ansatz in
\eqref{relalphas} one indeed reproduces the $\a_S$ of \cite{BH2},
while our result seems to suggest the need for $\cW$ corrections to
the mass. To test this possibility it would be desirable to compute
the shift in the mass in an independent way.\footnote{A modification
of the relation $M = M_{\text{BTZ}}$ was also proposed in \cite{mass} via a direct construction of the corresponding conserved charge in the CS formulation.} However, due to the modified asymptotic behaviour of the 
black hole solutions this is not straightforward: at infinity the
metric has the same radial dependence as an AdS space with half the
radius of the vacuum solution of our field equations \eqref{geq} and
\eqref{phieq}. The discrete jump in the asymptotic behaviour could
also create problems of convergence of the integrals involved in
Wald's proof of the first law of black hole thermodynamics. 
This issue deserves further study.

\section{Discussion}\label{sec:discussion}

We considered the minimal coupling of Fronsdal's action
to gravity in three space-time dimensions and we showed that at the lowest order in the higher-spin field it
preserves the same amount of gauge symmetry as the free action. The
resulting two-derivative coupling (that is not available in higher
space-time dimensions \cite{old-LC,Metsaev,MMR,ST,JLT,vertices-frame}) is
unique up to field redefinitions, and it does not require a
cosmological constant. In the spin-3 case we exhibited its relation
with the $SL(3,\mathbb{R}) \times SL(3,\mathbb{R})$ Chern-Simons
action that describes the coupling in a frame-like language. We also proved that a complete
metric-like action would require higher-order corrections in the
spin-3 field to preserve the gauge invariance, but the frame-like
formulation indicates that neither fields of different spin nor
higher-derivative couplings are necessary.

A natural extension of the present work would be to identify the full
metric-like counterpart of the $SL(3,\mathbb{R}) \times
SL(3,\mathbb{R})$ Chern-Simons action. The major simplification of the
spectrum in comparison to all known higher-spin gauge theories in
$D\geq 4$ gives hope to achieve this goal, although the next order
is already quite intricate and at present it is not even clear whether
the action has to be polynomial in the spin-3 field. The situation is as if in gravity we only knew the action up to some order in the graviton field.
Without an understanding of the geometric principles --- covariant derivatives, 
curvatures, etc.\ --- and of the full nonlinear diffeomorphism symmetry, 
this action would look mysterious. In the higher-spin case the ``geometric'' structures which are implied by the extension of diffeomorphism 
to include higher-spin gauge symmetries are unknown.  
Some attempts in this direction are reported e.g.\ in \cite{curvatures,DD,nonlocal,unconstr,alpha->nonlocal}, where reformulations of the free theory in terms of higher-spin curvatures were studied. 
In our setup progress in this direction might come from
abandoning metric compatibility and starting from a
Palatini-like description of the dynamics. 
The elimination of the auxiliary fields from the frame-like action, that 
we analysed in sec.\ \ref{sec:symm}, also gives hope that a simple way to group all non-linearities exists, but a detailed analysis of the next corrections is needed. 

Another interesting direction to
be explored is the extension of all previous considerations to
$SL(N,\mathbb{R}) \times SL(N,\mathbb{R})$ Chern-Simons theories and
to their $N \to \infty$ limits \cite{Blencowe1,Blencowe2}, possibly
including the matter couplings of \cite{VP} or generalisations thereof.

\section*{Acknowledgements}

We performed or checked various computations with $xAct$ packages for
Mathematica \cite{xAct}, and we are grateful to T.\ Nutma for his
advise and for the use of his package $xTras$. A.C., S.F.\ and
S.P.\ acknowledge the Erwin Schr{\"o}dinger Institute in Vienna for
hospitality, and the participants in the ESI workshop on Higher Spin
Gravity for feedback on a presentation of some of the results
contained in this paper. In particular, we would like to thank M.\ Ammon 
for a suggestion which provoked the analysis presented in sec.\ \ref{sec:wald}. 
We also thank M.\ Ba{\~n}ados, C.\ Eling, E.\ Joung, M.\ Henneaux, J.\
Mei, Y.\ Oz, M.\ Taronna and A.\ Virmani for discussions.

\begin{appendix}

\section{Conventions}\label{sec:conventions}
 
A pair of parentheses denotes the symmetrisation of the indices it
encloses with weight one, such that for instance
\begin{equation}
A_{(\mu}\, B_{\nu)} \, = \frac{1}{2} \left(A_{\mu}\, B_{\nu} +
A_{\nu}\,B_{\mu} \right) \ .
\end{equation}
In a similar fashion a pair of square brackets denotes the antisymmetrisation of
the indices it encloses. Note that this convention differs from the
one employed in our earlier publication~\cite{spin3}.
We adopt the mostly plus convention for the metric and  
our curvature conventions are
\begin{equation}
[\,\nabla_\m \,, \nabla_\n\,]\, \o_{\r} \,=\, R_{\m\n\r}{}^\s \o_\s \, , \qquad
R_{\m\n}\,=\, R_{\m\r\n}{}^\r \, .
\end{equation}
We often omit contracted indices in the traces of a tensor: for instance $\vf_\mu \equiv \vf_{\mu\l}{}^\l$.
 
The algebra $sl(3,\mathbb{R})$ can be given in terms of generators
$J_{a}$ and $T_{ab}$ with the commutation relations\footnote{With respect to \cite{spin3} we choose $\s = -1$.}
\begin{subequations}\label{sl3}
\begin{align}
& \left[\, J_a \comma J_b \,\right] \, = \, \e_{abc}\, J^c \, , \label{JJ} \\[6pt]
& \left[\, J_a \comma T_{bc} \,\right] \, = \, 2\,\e^d{}_{a(b} T_{c)d} \, , \label{JT} \\[6pt]
& \left[\, T_{ab} \comma T_{cd} \,\right] \, = \, -\, 2 \left(\, \h_{a(c} \e_{d)be} + \, \h_{b(c} \e_{d)ae} \,\right) J^e \, , \label{TT}
\end{align}
\end{subequations}
with $T_{[ab]} = \eta^{ab}\, T_{ab}=0$. The Levi-Civita symbol is
defined such that 
\begin{equation}
\e^{012} \, = \, - \ \e_{012} \, = \, 1 \, ,
\end{equation}
and indices are raised and lowered with
$\eta_{ab}=\text{diag} (-1,1,1)$. An explicit $3 \times 3$ matrix representation for the $T_{ab}$ is given by
\begin{equation}
T_{ab} = \left(J_a J_b+J_b J_a-\frac{2}{3}\, \eta_{ab}\, J_c J^c\right)\,,
\end{equation}
with $J_a$ in the three-dimensional representation of $sl(2,\mathbb{R}) \hookrightarrow sl(3,\mathbb{R})$. We denote by $\{J_{A}\}$ a set of five
independent generators built from the $T_{ab}$, and the set of all
generators $\{J_{a},J_{A} \}$ is denoted by $\{J_{\cA} \}$.

We normalise the Killing form to be one half of the matrix trace in
the fundamental representation of $sl(3,\mathbb{R})$,
\begin{equation}
\kappa_{\cA\cB} = \frac{1}{2}\, \tr \left(J_{\cA}\,J_{\cB} \right)\, ,
\end{equation}
such that e.g.\ $\kappa_{ab}=\eta_{ab}$ and $\k_{aB} = 0$. The anti-symmetric and
symmetric structure constants are defined as
\begin{align}
f_{\cA\cB\cC} &= \frac{1}{2}\, \tr \left([J_{\cA},J_{\cB}]J_{\cC}\right) \, ,\\
d_{\cA\cB\cC} &= \frac{1}{2}\, \tr \left(\{ J_{\cA},J_{\cB}\}
J_{\cC}\right) \, ,
\end{align}
such that $f_{abc}=\epsilon_{abc}$ and $d_{abc}=0$.
The structure constants satisfy  a number of identities, and in the main text we use
\begin{subequations}
\begin{align}
f_{a}{}^{bc}\,f_{bc}{}^{d} \,& =\, -2\,\delta_{a}^{d} \ ,\\
f_{ab}{}^{c} \,f_{c}{}^{de} \,& =\, -
\left(\delta_{a}^{d}\,\delta_{b}^{e}-\delta_{a}^{e}\,\delta_{b}^{d}
\right) \ ,\\
d_{Abc}\,\kappa^{bc} \,&=\, 0 \ ,\\
d_{Abc}\, d^{A}{}_{de} \,&=\, -\,\tfrac{2}{3}\,\kappa_{bc}\, \kappa_{de} 
+ 2\,\kappa_{d(b}\, \kappa_{c)e} \ .\label{app:dd-identity}
\end{align}
\end{subequations}

\section{Spin-3 transformation of the metric}\label{sec:spin3}

In this appendix we want to show that the spin-3 transformation of the
metric in~\eqref{varmetricfromframe} derived from the frame-like
formulation (which we call $\delta^{I} g_{\mu\nu}$ in the following)
can be expressed in terms of metric $g$ and spin-3 field $\vf$ as
in~\eqref{dg-x} (which we will call
$\delta^{II}g_{\mu\nu}$).

We first expand the $sl(3)$-covariant derivative $\cD$
in~\eqref{varmetricfromframe} as the sum of the $sl(2)$-covariant
derivative $\os{0}{\cD}$ and the non-$sl(2)$ part $\Omega$ of the spin
connection,
\begin{equation}
\delta^{I}g_{\mu\nu}=\, -\,6\,d_{Abc}\, e^{b}_{\rho}\, e^{c}_{\sigma}\,
\xi^{\rho\sigma}\, \left(\os{0}{\cD}{}_{(\mu}^{\phantom{A}}E_{\nu)}^{A} 
+f^{A}{}_{Bd}\, \Omega^{B}_{(\mu}\,e^{d}_{\nu)}\right) + \, \mathcal{O}
\!\left(E^{3}\right) \ , 
\end{equation}
and replace $\Omega$ by the expression~\eqref{spin3connection} to
arrive at
\begin{align}
\delta^{I}g_{\mu\nu}&=\, -\, \xi^{\rho\sigma}\, e^{e}_{\rho}\,
e^{f}_{\sigma}\,e^{a}_{(\mu}\,e^{b}_{\nu)}\,
e^{m}_{\alpha}\,e^{n}_{\gamma}\,g^{\alpha\beta}\,g^{\gamma\delta}\,
\os{0}{\cD}{}_{\beta}E_{\delta}^{A} \nonumber\\
& \times \left(6\kappa_{m(a}\,\kappa_{b)n}\,d_{Aef}
+d^{D}{}_{ef}\,f_{mn}{}^{d}\,f_{CD(a}
\left(\kappa_{b)d}\,\delta_{A}^{C} 
+ f_{b)d}{}^{g}\,f_{Ag}{}^{C} 
+f_{b)E}{}^{C}\,f_{Ad}{}^{E}\right)
 \right)\ .
\label{expr1}
\end{align}
On the other hand we can start from the
expression~\eqref{dg-x} and expand it in vielbeins. For
that we first augment the Levi-Civita covariant derivative
$\nabla_{\!\mu}$ to the $sl(2)$-covariant derivative $\os{0}{\cD}{}_{\mu}$
because then it acts trivially on the $sl(2)$-components $e^{a}$ of
the vielbein (up to $\mathcal{O} (E^{2})$ corrections). We obtain
\begin{align}
\delta^{II}g_{\mu\nu}&=\, \xi^{\rho\sigma}\, e^{e}_{\rho}\,
e^{f}_{\sigma}\,e^{a}_{(\mu}\,e^{b}_{\nu)}\,
e^{m}_{\alpha}\,e^{n}_{\gamma}\,g^{\alpha\beta}\,g^{\gamma\delta}\,
\os{0}{\cD}{}_{\beta}E_{\delta}^{A} \nonumber\\
& \times \Big( 4\kappa_{n(a}\,d_{b)A(e}\,\kappa_{f)m}
+ 2d_{Aab}\,\kappa_{m(e}\,\kappa_{f)n}
-4 d_{Aef}\,\kappa_{m(a}\,\kappa_{b)n}
-8 \kappa_{m(a}\,d_{b)A(e}\,\kappa_{f)n}\nonumber\\
&\qquad 
+ 4d_{Am(e}\,\kappa_{f)(a}\,\kappa_{b)n}
+ 4d_{Am(a}\,\kappa_{b)(e}\,\kappa_{f)n}
+ 4d_{A(a|(e}\,\kappa_{f)|b)}\,\kappa_{mn}
- 8d_{An(a}\,\kappa_{b)(e}\,\kappa_{f)m}\nonumber\\
&\qquad 
-8 d_{An(e}\,\kappa_{f)(a}\,\kappa_{b)m}
+2d_{Amn}\,\kappa_{a(e}\,\kappa_{f)b}
-2d_{Aef}\,\kappa_{ab}\,\kappa_{mn}
-4d_{Am(e}\,\kappa_{f)n}\,\kappa_{ab}\nonumber\\
&\qquad 
+8 d_{An(e}\,\kappa_{f)m}\,\kappa_{ab}
-2d_{Amn}\,\kappa_{ab}\,\kappa_{ef}
-2d_{Aab}\,\kappa_{ef}\,\kappa_{mn}
-4d_{Am(a}\,\kappa_{b)n}\,\kappa_{ef}\nonumber\\
& \qquad 
+8d_{An(a}\,\kappa_{b)m}\,\kappa_{ef}
\Big) \ .
\label{expr2}
\end{align}
Identifying the two expressions~\eqref{expr1} and~\eqref{expr2} then
amounts to checking an identity for the structure constants which can
easily be done with the help of a computer.

\section{Tensors of arbitrary rank}\label{sec:spin}

One can extend the considerations of sec.\ \ref{sec:coupling} to
symmetric tensors of arbitrary rank. To this end it is convenient to switch to a more compact notation: in this section repeated covariant (or
contravariant) indices denote a symmetrisation. Moreover, symmetrised indices belonging to the same tensor are substituted by a single Greek letter with a label counting the total number of indices. For instance, the covariantised Fronsdal tensor can be written as
\begin{equation}
\cF_{\m_s} =\, \Box\, \vf_{\m_s} -\,\frac{s}{2} \left(\, \nabla^\l  \nabla_{\!\m}\, \vf_{\m_{s-1}\l} 
+ \nabla_{\!\m} \nabla^\l\,  \vf_{\m_{s-1}\l} \,\right)+  \binom{s}{2}\, \nabla_{\!\m} \nabla_{\!\m}\,
{\vf}_{\,\m_{s-2}} \, ,
\end{equation}
where $\vf_{\mu_{s-2}}=g^{\rho\sigma}\vf_{\mu_{s-2}\rho\sigma}$.
With these conventions its variation under
\begin{equation} \label{gen-fronsdal}
\d \vf_{\m_s} =\, s\,\nabla_{\!\m}\, \x_{\,\m_{s-1}} \, , \qquad \textrm{with} \quad {\x}_{\,\m_{s-3}} \,=\, 0
\end{equation}
takes essentially the same form as in the spin-3 case,
\begin{align}
\d \cF_{\m_s} =\, & - 6\binom{s}{3} \x^{\a\b}{}_{\m_{s-3}} \nabla_{\!\m}\, R_{\,\a\m\,\m\b}
- 9 \binom{s}{3} R_{\,\a\m\,\m\b} \nabla_{\!\m}\,
\x_{\m_{s-3}}{}^{\a\b} + 2 \binom{s}{2}
R_{\,\a\m\,\m\b} \nabla^\a\, \x_{\m_{s-2}}{}^\b \nonumber\\
& - 2 \binom{s}{2} \x^\a{}_{\m_{s-2}} \nabla_{\!\a}\, R_{\m\m} + \frac{s}{2}\, R_{\m\a}
\nabla^\a\, \x_{\m_{s-1}} -3 \binom{s}{2} R_{\m\a} \nabla_{\!\m}\,
\x_{\m_{s-2}}{}^\a \, .
\end{align}
In Fronsdal's approach the double trace of the fields is forced to
vanish. As a result, at the lowest order in $\vf$ the spin-3 example
already captures all features of the general case because one cannot
construct other curvature terms than those in \eqref{LF}. The most
general Lagrangian that is quadratic in $\vf$ and contains at most two
derivatives is
\begin{align}
\cL_F \, & = \frac{\sqrt{-g}}{16\p G} \, \bigg\{\,\vf^{\,\m_s} \left( \cF_{\m_s} 
- \12 \binom{s}{2} g_{\m\m}\,{\cF}_{\m_{s-2}} \right) + \frac{1}{\ell^2} \left(\, m_1\, \vf_{\m_{s}} \vf^{\,\m_{s}} \,
+\, m_2\,{\vf}_{\m_{s-2}}\, {\vf}^{\,\m_{s-2}} \,\right)\nonumber\\
& +\, \binom{s}{2}\, R_{\a\b} \left(\, 
k_1\,  \vf^{\,\a}{}_{\m_{s-1}}\, \vf^{\,\b\,\m_{s-1}} 
+ k_2\, \vf^{\,\a\b}{}_{\m_{s-2}}\, \vf^{\,\m_{s-2}} 
+ (s-2)\,k_3\, {\vf}^{\,\a}{}_{\m_{s-3}}\, {\vf}^{\,\b\,\m_{s-3}} \,\right) \nn\\ 
& +\, \binom{s}{2}\, R \left(\, k_4\, \vf_{\m_{s}} \vf^{\,\m_{s}} \,
+\, k_5\, {\vf}_{\m_{s-2}}\, {\vf}^{\,\m_{s-2}}
\,\right) \bigg\} \, .
\label{lag}
\end{align}
Requiring that on (A)dS the terms in $\ell^{-2}$ build the Fronsdal mass term \cite{Fronsdal-AdS} then implies 
\begin{eqnarray}
m_1 \!&=&\! s \left\{ (s-1)(k_1+3k_4) - \frac{3s-5}{2} \right\} \, , \\
m_2 \!&=&\! s(s-1) \left\{ k_2 + (s-2)k_3 + 3k_5 + \frac{s(3s-1)-6}{8} \right\} \, .
\end{eqnarray}
If these conditions are satisfied, the Lagrangian $\cL = \cL_{EH} +
\cL_F$ is invariant, up to linear order in $\vf$ and up to total
derivatives, under the simultaneous transformations
\eqref{gen-fronsdal} and
\begin{align}
\d g_{\m\m} \, & =\, \binom{s}{2} \, \Big\{\, a_1\, \vf_{\m\m\,\a_{s-2}} \nabla\cdot \x^{\a_{s-2}} \,
+\, (a_2+b_{2})\, \vf_{\m\,\a_{s-1}} \nabla_{\!\m}\, \x^{\a_{s-1}} \,
+\, a_3\, \vf_{\m\,\a\,\b_{s-2}} \nabla^\a \x_\m{}^{\b_{s-2}} \nn \\
& \!\!\!\! +\, a_4\,{\vf}_{\m\,\a_{s-3}} \nabla\cdot \x_\m{}^{\a_{s-3}} \,
+\, (a_5+b_{5})\,{\vf}_{\a_{s-2}} \nabla_{\!\m}\, \x_\m{}^{\a_{s-2}} \,
+\, a_6\,{\vf}_{\a\,\b_{s-3}} \nabla^\a \x_{\m\m}{}^{\b_{s-3}} \nn \\[6pt]
& \!\!\!\! +\, b_1\, \x^{\a\,\b_{s-2}} \nabla_{\!\a}\, \vf_{\m\m\,\b_{s-2}} \,
+\, b_2\, \x^{\a_{s-1}} \nabla_{\!\m}\, \vf_{\m\,\a_{s-1}} \,
+\, b_3\, \x_\m{}^{\a_{s-2}} \nabla\cdot\vf_{\m\,\a_{s-2}}  \label{gauge-s} \\[6pt]
& \!\!\!\! +\, b_4\, \x_\m{}^{\a\,\b_{s-3}} \nabla_{\!\a}\,{\vf}_{\m\,\b_{s-3}} \,
+\, b_5\, \x_\m{}^{\a_{s-2}} \nabla_{\!\m}\,{\vf}_{\a_{s-2}} \,
+\, b_6\, \x_{\m\m}{}^{\a_{s-3}} \nabla\cdot{\vf}_{\a_{s-3}} \nn \\[1pt]
& \!\!\!\!+ \,g_{\m\m} \left( c_1\, \vf_{\a\,\b_{s-1}} \nabla^\a \x^{\b_{s-1}} 
+ \,c_2\,{\vf}_{\a_{s-2}} \nabla\!\cdot \x^{\a_{s-2}} 
+ \,d_1\, \x^{\a_{s-1}} \nabla\!\cdot \vf_{\a_{s-1}} 
+ \,d_2\, \x^{\a\,\b_{s-2}} \nabla_{\!\a}{\vf}_{\b_{s-2}} \right)\!\! \Big\} . \nn
\end{align}
As in the spin-3 case, all coefficients in \eqref{gauge-s} are fixed
except $b_{2}$ and $b_{5}$, which parameterise field-dependent
diffeomorphisms.

The coefficients $a_{i}$ and $c_{i}$, which multiply terms with the derivative
acting on the gauge parameter, depend on the $k_i$, 
\begin{alignat}{3}
& a_1 \,=\, 2 k_2 \, , \qquad & & a_2 \,=\, \frac{4s + 2(s-1)k_1 -
2}{s-1}\, , \nn \\
& a_3 \,=\, 2(s-1)(k_1-2) + 2 \, , \qquad & &  a_4 \,=\, (s-2) (2s +
4k_3 - 3)\, , \nn\\[6pt] 
& a_5 \,=\, 2 ( 4s + k_2 - 8 ) \, , \qquad & & 
a_6 \,=\, (s-2)k_2 \, , \nn \\
& c_1 \,=\, -\, \frac{2s(s-1)(k_1+2k_4-2)+2s^2}{s-1} \, , \quad & & \nn \\ 
& c_2 \,=\, -\, 4\, (k_2 + (s-2)k_3 + 2k_5) - s(s-2) \, . \quad & & 
\end{alignat}
The remaining coefficients do not depend on the free parameters in the
Lagrangian:
\begin{alignat}{5}
& b_1 \,=\, 4 \, , \qquad\quad & & b_3 \,=\, 8(s-2) \, , \qquad\quad & & b_4 \,=\, -\,4(s-1)(s-2) \, , \nn \\[5pt]
& b_6 \,=\, 2 (s-2) \, , \qquad\quad & & d_1 \,=\, -\, 4(s-2) \, , \qquad\quad & & d_2 \,=\, 2(s-2)(s+1) \, .
\end{alignat}
As in the spin-3 case one can set to zero all terms with the derivative acting on
$\xi$ by choosing $b_{2}=-a_{2}$, $b_{5}=-a_{5}$, and
\begin{equation}
k_1 = \frac{2s-3}{s-1} \,, \quad\ k_2 = 0 \,, \quad\ k_3 = \frac{3-2s}{4} \,,
\quad\ k_4 = -\,\frac{1}{2} \,, \quad\ k_5 = \frac{(s-2)(s-3)}{8} \,.
\end{equation}
The coupling is again unique since one can remove all contributions in the Ricci tensor with the field redefinition
\begin{equation}
\begin{split}
g^{\text{(new)}}_{\m\m} =\, g_{\m\m} 
\,& -\, \binom{s}{2} \bigg\{\, k_1\, \vf_{\m\,\a_{s-1}}\, \vf_{\m}{}^{\a_{s-1}} 
\,+\, k_2\, \vf_{\m\m\,\a_{s-2}}\, {\vf}^{\,\a_{s-2}} 
\,+\, k_3\, {\vf}_{\m\,\a_{s-3}}\, {\vf}_{\m}{}^{\a_{s-3}} \\
& -\, g_{\m\m} \left[\, (k_1+2k_4)\, \vf_{\a_s}\, \vf^{\,\a_s} 
\,+\, (k_2+k_3+2k_5)\,{\vf}_{\a_{s-2}}\,{\vf}^{\,\a_{s-2}} \,\right] \bigg\} \, .
\end{split}
\end{equation}

Note that the terms in \eqref{lag} should appear in the metric-like
counterpart of any Chern-Simons theory involving $\vf_{\m_1 \cdots\,
\m_s}$ since they provide a diffeomorphism invariant version of
Fronsdal's kinetic operator. On the other hand, we expect inequivalent
higher-order completions of \eqref{lag}, corresponding to different
Chern-Simons theories. The simplest examples of this type involve
various higher-spin fields, and it is not clear whether for $s > 4$
the gauge symmetry of \eqref{lag} can also be preserved at higher
orders without considering at the same time symmetric tensors of
different rank (see also \cite{frame-3d} for a direct construction of
higher-spin interactions in the frame-like formulation).

\end{appendix}


\end{document}